\documentclass[12pt]{article}

\usepackage{amsmath,amsfonts,amssymb}
\usepackage{float}

\usepackage{graphicx}
\newcommand{\mathsym}[1]{{}}
\def\id{\protect{{1 \kern-.28em {\rm l}}}}

\def\be{\begin{eqnarray}}
\def\ee{\end{eqnarray}}
\def\ba{\begin{eqnarray}}
\def\ea{\end{eqnarray}}
\def\p{{\partial}}

\makeatletter
\renewcommand\section{\@startsection {section}{1}{\z@}%
                                   {-3.5ex \@plus -1ex \@minus -.2ex}%
                                   {2.3ex \@plus.2ex}%
                                   {\normalfont\large\bfseries}}
\renewcommand\subsection{\@startsection{subsection}{2}{\z@}%
                                   {-3.25ex\@plus -1ex \@minus -.2ex}%
                                   {1.5ex \@plus .2ex}%
                                   {\normalfont\normalsize\bfseries}}

\makeatother

\def \foot {\footnote}
\def \bi{\bibitem}
\def \tr {{\rm tr}}
\def \ha {{1 \over 2}}
\def \td {\tilde}
\def \ci{\cite}

\def\S{{\mathcal S} }

\def \E {{\mathcal  E}} \def \J {{\mathcal  J}}

\def\p{\phi}

\def\g{\gamma}
\def\ov{\over}

\def\J{{\mathcal J}}

\def\E{{\mathcal E}}

\def\l{\lambda}

\def\foot{\footnote}

\def \ci {\cite}

\def \fo { {1\ov 4}}

\def \l  {\lambda}

\def \S {{\rm S}}

\def \td {\tilde}

\def \D {\Delta}

\def \bi{\bibitem}
\def \la {\label}

\def \l {\lambda}
\def\foot{\footnote}

\def \adss {$AdS_5 \times S^5~$ }
\newcommand{\rf}[1]{(\ref{#1})}
\def \ov {\over}

\def\cc{\circ}

\def \ha{{1\ov 2}}

\def \no {\nonumber}

\def \J {\mathcal{J}}

\def \E {{\cal E}}

\def \S {{\cal S}}
\def \J {{\cal J}}

 \def \bb {\bar \beta}

\def \bi{\bibitem}
\def \la {\label}

\def\foot{\footnote}

\def \sql {{\sqrt \lambda}}

\def \adss {$AdS_5 \times S^5$\ }

\def \D {\Delta}

 \def \p {\phi}

\def \ov {\over}

\def \varpi {{\rm w}}
\def \OO {{\cal O}}

\def \te {\theta}

\def \S  {{\rm S}}

\def \C {{\cal C}}

 \def \J {{\cal J}}
 \def \S {{\cal S}}
 \def \E {{\cal E}}

\def \p {\phi}

\def \os  {\OO({\textstyle{ 1\ov \sqrt{\lambda}}})}

 \def \sql {\sqrt{\lambda}}

\def \cc {{c }} 
\def \OO {{\cal O}}
\def \te {\textstyle}
\def \fl {\sqrt[4]{\lambda}}

\def \fo {{\textstyle{1 \ov4}}}
\def \rx {{\rm x}}
\def \hg {{\hat \g}}

\def \C  {{\rm C}}
\def \hC  {{\rm \hat  C}}
\def \dd  {{\rm d}}
\def \bb {{\rm b}}
\def \dDelta {2}
\def \sql {{\sqrt{\lambda}}}
\def \ed {\end{document}}
 \def \an {{\rm an}} \def \nan {{\rm nan}}

\def \td {\tilde}
 
\newcommand{\mc}{\mathcal }
  
\begin{document}


\overfullrule=0pt
\parskip=2pt
\parindent=12pt
\headheight=0in \headsep=0in \topmargin=0in \oddsidemargin=0in

\vspace{ -3cm}
\thispagestyle{empty}
\vspace{-1cm}

\rightline{Imperial-TP-AAT-2012-04}


\begin{center}
\vspace{1cm}
{\Large\bf  
More about ``short''  spinning quantum strings 
\vspace{1.2cm}
}

\vspace{.2cm}
 {M. Beccaria$^{a}$, 
 and  A.A. Tseytlin$^{b,}$\footnote{Also at Lebedev  Institute, Moscow. }
}

\vskip 0.6cm

{\em 
$^{a}$Dipartimento di Matematica e Fisica ``Ennio De Giorgi'', \\ Universita' del Salento \& INFN, 
                     Via Arnesano, 73100 Lecce, Italy\\
\vskip 0.08cm
\vskip 0.08cm 
$^{b}$Blackett Laboratory, Imperial College,
London SW7 2AZ, U.K.
 }

\vspace{.2cm}

\end{center}

\begin{abstract}
 
We continue investigation of the  spectrum of  semiclassical
 quantum strings in $AdS_5 \times S^5$  on the examples 
 of folded  $(S,J)$ string   (with spin $S$ in $AdS_5$  and orbital momentum $J$ in $S^5$)
dual to an $\mathfrak{sl}(2)$  sector state  in gauge theory 
 and its $(J',J)$  counterpart  with spin $J'$ in $S^5$ dual to an $\mathfrak{su}(2)$  sector state.
We study the limits  of small spins and large $J$ at weak and strong coupling,
 pointing out that terms linear in spins  provide a generalization of 
 ``protected''  coefficients  in the energy that are given 
 by finite polynomials in 't Hooft  coupling $\lambda$ (or square of string tension)
   for any value of $\lambda$. We  propose an  expression for the coefficient of the 
term linear in spin $J'$  in the $(J',J)$ string 
energy  which should be  the $\mathfrak{su}(2)$  sector 
counterpart of the ``slope function'' in the  $\mathfrak{sl}(2)$  sector 
suggested   by Basso in arXiv:1109.3154.

\end{abstract}

\allowdisplaybreaks

\newpage
\setcounter{equation}{0} 
\setcounter{footnote}{0}
\setcounter{section}{0}
\renewcommand{\theequation}{1.\arabic{equation}}
 \setcounter{equation}{0}
\setcounter{equation}{0} \setcounter{footnote}{0}
\setcounter{section}{0}

\def \os {O(\textstyle{ {1 \ov (\sqrt{\lambda})^2}} )}
\def \ost {O(\textstyle{ {1 \ov (\sqrt{\lambda})^3}} )}
\def \cc {{c }} 
\def \OO {{\cal O}}
\def \te {\textstyle}
\def \fl {\sqrt[4]{\l}}

\def \ha {{{\textstyle{1 \ov2}}}}
\def \fo {{\textstyle{1 \ov4}}}
\def \rx {{\rm x}}
\def \hg {{\hat \g}}

\def \C  {{\rm C}}
\def \hC  {{\rm \hat  C}}
\def \dd  {{\rm d}}
\def \bb {{\rm b}}
\def \dDelta {2}
\def \sql {{\sqrt{\l}}}

 \def \an {{\rm an}} \def \nan {{\rm nan}}
 \def \nm {\tilde n_{11}}
 \def \tn {{\tilde n}}  \def \uni {{\rm inv}}
\def \ttn  {{\bar n}_{11}}
\def \ed {\end{document}}

\def \sj {\te { S\ov J}} 
\def \sjj   {\te { \S\ov \J}}  

\def \ca {{\rm a}} \def \cb {{\rm b}} \def \cc {{a}}  \def \ct {{\rm  v }}  
\def \te {\textstyle} \def \cm {{\rm b}}

\def \ccc {{\tilde a}}

\def \v {h} \def \u {\td h} 
\def \mksl {\mathfrak{sl}(2)}
\def \mksu {\mathfrak{su}(2)}
\def \ccm {{\bar b}}
\tableofcontents

\section{Introduction and summary}

In this paper we continue investigation of the spectrum of  semiclassical
 quantum strings in 
\adss  in the small-spin  limit  (``short'' strings) 
 \ci{tt,rt1,gssv,rt2,bm,gva,bgmrt}.
We shall clarify the structure of the expansion of the string energy $E$
 in   orbital momentum $J$ in $S^5$.
 In particular, we  shall  compare the  energies of folded  $(S,J)$ string 
 in $AdS_3 \times S^1$ 
and folded $(J',J)$ string in $R \times S^3$ 
(representing gauge-theory states  in the $\mksl$ and
 $\mksu$ sectors respectively).
 We will be interested  in the small 
$\S$  (or small $\J'$) expansion of the  energy at fixed $S^5$  orbital momentum $\J$, 
i.e. in the limit $ {\S \ov \J } \ll 1$ (or  $ {\J' \ov \J } \ll 1$).\foot{We 
   shall use the following notation:
the string tension is $T= {\sqrt{\lambda} \ov 2 \pi}$,  the semiclassical parameters  are 
$\S_i=  { S_i \ov  \sqrt{\lambda}}, \ \J_i =   { J_i \ov  \sqrt{\lambda}}$. 
 The spin in $S^5$  will be denoted as 
 $ J'\equiv  J_1 $ and the orbital momentum as 
 $J\equiv  J_2$.}
 We will  suggest the   expression for the $J'$
``slope''  function 
  which is the $\mksu$  sector
   counterpart of the expression  of \ci{bas} in the $\mksl$ sector.

Let  us first  discuss
  the general structure  of the energy $E$ of a semiclassical  string state 
and  compare it with the corresponding expression for the gauge-theory   dimension
$\D$. 
For definiteness, let  us  consider string states with 
  spin $S$ in $AdS_5$ and    orbital momentum $J$ in $S^5$
  dual to 
 gauge theory  states  from the 
 $\mksl$ sector  represented  by operators  like     $\tr ( D_+^S \Phi^J)$  
(similar  discussion will apply  to states from the $\mksu$ sector). 
In perturbative planar  gauge theory    one first  expands $\D=E$ 
in  $ \lambda \ll 1$  for fixed spin $S$  and $J$ 
\be E\equiv \Delta =  S +  J +   \g (S,J,  \lambda)\  ,  \ \ \ \ \ \quad 
\g=\sum^\infty_{n=1}  \lambda^n \g_n (S,J)   \ .    \la{dima} \ee
One  may then further expand $\g_n$  in $S$ and $J$, 
 e.g.,    in large    $J$  for fixed $S$. 
The  semiclassical  string theory limit  corresponds to 
first taking  $ \sqrt{\lambda}  \gg  1 $
for fixed   semiclassical  parameters  
$\S= {S \ov \sqrt{\lambda}}  , \   \J = { J \ov \sqrt{\lambda}}  $ 
 (which means  that $S$ are $J$ are assumed to be 
as large as $\sqrt{\lambda}$) 
\be  E = J +S  +  e (\S,\J,  \sqrt{\lambda})  \ , \ \ \ \ \ \ \quad 
e= \sum^\infty_{p=0} {1 \ov (\sqrt{\lambda})^{p-1}  }  e_p (\S,\J)  
  \ ,   \la{stri} \ee
 and may then further  expand
   $e_p$  for large or small  $\S,\J$.
   
 The  AdS/CFT  duality implies that the 
  final expression for $E$ in  \rf{dima}  summed up in $\lambda$ and then 
 expanded at strong coupling (i.e. in $ {1 \ov \sqrt{\lambda}} \ll 1 $ ) should match 
 \rf{stri}, i.e. $\g (S,J,  \lambda)  =  e (\S,\J,  \sqrt{\lambda}) $,
  but  the  two expansions are a priori  very different  and cannot be 
   compared directly. 
   Still, as 
 was  noticed  starting with \ci{bmn,ft2,bmsz,ft3,kru},  it is possible to
establish a more direct relation between the perturbative 
gauge theory and string theory 
results  for the few leading  terms in the above expansions 
 by considering large charge  limits in which supersymmetry protection
 effectively comes into play.

Let  us start with gauge theory and assume that $J \gg 1$ while $S$ is fixed. 
Ignoring wrapping  corrections which  should  be exponentially suppressed 
at large $J$,   the corresponding dimension should be described by the Asymptotic Bethe 
Ansatz (ABA)  \ci{bes}.  If one  formally ignores the contribution of the dressing 
 phase \ci{afs,bt,hl,bhl}   and 
starts with  the original  BDS 
 Bethe Ansatz  \cite{bds} then  one finds that the $1/J$ expansion of 
 $\g_n$ in \rf{dima} has 
 the following structure 
    ($n=1,2,3,..$)\foot{The leading terms in this  expansion 
    of $E$  are consistent with the BMN scaling, i.e. 
    depend on $J$ through the combination ${\lambda \ov J^2}$.} 
\be \la{dma}
 \g^{(0)}_n = { 1 \ov  J^{2n-1} }  \sum^{\infty}_{k=1}  { \ca_{nk} (S)  \ov J^k } 
\,, \qquad \qquad
\ca_{nk} (S) =   \sum^{k  }_{m=1}  \ca_{nk;m}  S^m \ , 
  \ee
where 
\be \la{ex}
 \ca_{n1} =  \ca_{n1;1} S \ , \ \ \  \ca_{n2} =  \ca_{n2;1} S +   \ca_{n2;2} S^2   \ , \ \ \ 
 \ca_{n3} =  \ca_{n3;1} S +   \ca_{n3;2} S^2 +   \ca_{n3;3} S^3    \ , \ \ ... \  \ .  \ee
This  large $J$ expansion   may be rewritten also as  
\be \la{ma}
 \g^{(0)}_n = { 1 \ov  J^{2n-1} }  \sum^{\infty}_{k=0}  { \cc_{nk} ({S\ov J})  \ov J^k } 
\,, \qquad \qquad
\cc_{nk} ({\te{S\ov J}}) =   \sum^{\infty   }_{m=1}  \cc_{nk;m}  ({\te{ S\ov J}})^m \ , \ \ \ \ \
 \cc_{nk;m} =  \ca_{n,k+m;m} \ . 
   \ee
While   the functions $ \ca_{nk} (S)$  in \rf{dma} are finite 
polynomials in $S$, the 
 functions $ \cc_{nk} ({\te{S\ov J}}) $ are given by infinite
  series.\foot{Below we shall always assume that 
$u= { S \ov J}  < 1$  so that the series formally converges.} 
Eq.(\ref{dma}) implies  that each   $1/J^{k}$ term 
in $\gamma$
 receives contributions from a finite number of loop  orders only, thus 
excluding a possibility of a 
non-trivial ``interpolating'' functions of $\lambda$  as coefficients. 
In fact, the scaling (\ref{dma})  happens to  be 
 broken  starting with 4 loops ($n=4$) 
by the dressing phase contribution \ci{bes} that  leads   to additional contributions to
 $\g_n$  with $n \geq 4$:
 \be 
&&\g_{1,2,3}= \g^{(0)}_{1,2,3} \ , 
\ \ \ \ \ \ \ \ \ \  \ \ \ \ \ \ \ \ \ \ \  \ \ \ \ 
 \g_{n\geq 4}= \g^{(0)}_n + \g^{(1)}_n \ , 
  \la{h} \\
&& \g^{(1)}_n 
= { 1 \ov  J^{5} }  \sum^{\infty}_{k=0}  { \ccc_{nk} ({S\ov J})  \ov J^k } \ , 
\qquad \qquad
\ccc_{nk} ({\te{S\ov J}}) =   \sum^{\infty   }_{m=2}  \ccc_{nk;m}  ({\te{ S\ov J}})^m   \ , 
 \la{hi} \ \ \ \ \ \  n \geq 4 \ . 
 \ee
The contributions  due to the presence of the phase 
  producing the correction $\g^{(1)}_n $ appear to start  only with $S^2$ terms, i.e. 
  they do not influence 
  terms  linear in  $S$ which  determine  the slope function $h_1$ in \rf{sl}
below 
 \ci{bas,kol,bas2}. Indeed, as we shall explicitly demonstrate 
 in Appendix C, the first  non-trivial 4-loop   contribution of the phase
 leads to 
 \be 
  \g^{(1)}_4 
= \ccc_{40;2} { S^2 \ov  J^{7} }  + ...\ , \ \ \ \ \ \ \ \ \ \ \ \ \ \ \ 
\ccc_{40;2}= - {\zeta (3) \ov 32 \pi^2}   \ .  \la{k} \ee
The corresponding limit  on the semiclassical   string theory
side is  $\J \gg 1$ 
where $e_p$ in \rf{stri}  have the 
following structure
\be
&&e_{0,1,2,3,4} =e'_{0,1,2,3,4} \ , \ \ \ \ \ \ \ \ \   \ \  e_{p \geq 5}
=e'_p + e''_p \ , \la{ee} \\ 
&& \la{mai}
 e'_p = { 1 \ov \J^{p+1}  }  \sum_{q=1}^\infty  { \cm_{pq   }  (\S) \ov \J^q}  \ ,  
 \ \ \ \ \ \ \ \ 
  \cm_{pq   }  (\S) =  \sum^{q  }_{r=1}  \cm_{pq;r}  \S^r \ , \la{ch}
\ee
where 
$e''_p$   should have   ``softer'' $ { 1 \ov \J^{k}  }$  ($k < p+1$) prefactor 
than $ e'_p $ 
and is expected to  start   with $p=5$,\ \ \   $e''_5 \sim {\S^2 \ov \J^7}$.
While  both  $ e'_p $   and  $ e''_p $ receive  contributions
 from the quantum  part of the 
dressing phase \ci{bt,hl} in the ABA  (complementing the leading 
  ``classical'' AFS part \ci{afs}),  which start with 
  $1\ov \J^5$ terms in the 1-loop $e_1$, 
    $ e''_p $  would be  absent if one would ignore the quantum part of the 
    phase in 
    strong-coupling ABA. 
    
For example, $e_0$   entering the  classical  string energy with fixed $\S$ 
 may be written as
\be 
 e_0 =  { 1 \ov \J   } \Big( { \cm_{01;1}  \S\ov \J}  +  {   \cm_{02;1}  \S + \cm_{02;2} \S^2   \ov \J^2} 
 +  {   \cm_{03;1}  \S + \cm_{03;2} \S^2   +  \cm_{03;3} \S^3   \ov \J^3}  + ...\Big)  \ . \ee
The expansion of $e'_p$ in \rf{mai}  may  be  reorganized as\foot{The 
$ { 1 \ov \J^{p+1}  }$ scaling of $e'_p$ in \rf{mai}  translates  into 
$ { 1 \ov (\sqrt{\lambda})^{p-1} } { 1 \ov \J^{p+1}  }b_{p0}  ({\te {\S\ov \J}})
= { \lambda \ov J^{p+1}} b_{p0}  ({\te {S\ov J}})$
leading contribution to $e$ in \rf{stri}.} 
\be 
e'_p = { 1 \ov \J^{p+1}  }  
\sum_{q=0}^\infty { b_{pq}  ({\te {\S\ov \J}}) \ov \J^q}  \ ,  \ \ \ \ \ \ \ \ 
  b_{pq   }   ({\te{\S\ov \J}})  =  \sum^{\infty }_{r=1}  b_{pq;r}   ({\te {\S\ov \J}}) ^r \ , \ \ \ \ \ 
 b_{pq;r} =  \cm_{p,q+r;r}  \ . \la{cv}
\ee
Like $\ca_{nk} (S)$   and  $\cc_{nk} ({\te{S\ov J}}) $    in \rf{dma},\rf{ma}, 
 while  
$ \cm_{pq   }  (\S)$ in \rf{mai} are finite polynomials, the functions 
$ b_{pq   }   ({\te{\S\ov \J}})$   are given by infinite series. 
Explicitly,  for $e_p=e'_p$  with $p=0,1,2,3,4$ one has 
 \ci{ft3,bt} 
\be \la{ai}
&&
 e_0 = { 1 \ov \J }  \Big(b_{00} + { b_{02} \ov \J^2} +   
 { b_{04} \ov \J^4 }  + ... \Big)   \ ,
\   \ \ \ \  \  e_1 = { 1 \ov \J^2 }  \Big(b_{10} + { b_{12} \ov \J^2} +  
 { b_{13} \ov \J^3 }  + ... \Big)   \la{ki} ,\\
&&    e_2 = { 1 \ov \J^3 }  \Big(b_{20}  +  { b_{22} \ov \J^2}+   ...  \Big) \ ,
 \ \ \   \  
e_3 = { 1 \ov \J^4 }  \Big(b_{30}  + { b_{31} \ov \J }+ ... \Big) \ , \ \ \ \ 
e_4 = { 1 \ov \J^5 }  \Big(b_{40} + ... \Big) \la{aki}
 \ee 
For $e''_p$  we expects to find that 
\be 
e''_p = { 1 \ov \J^{5}  }  \ccm_{p0}({\te {\S\ov \J}}) + O\big({ 1 \ov \J^{6}  }\big) \ , \ \ \
\ \ \ \ \  \ccm_{p0}= \ccm_{p0;2} {\S^2\ov \J^2}  +  O\big({ {\S^3\ov \J^3}}\big)
\ , \ \ \ \ p \geq 5  \ . \la{y}
\ee
 An  advantage   of  organizing the expansion in terms of functions  
  of the ratio
\be  u= {S \ov J }= { \S \ov \J} \la{uuu}
\ee   is that it does  not explicitly 
 depend on the coupling $\lambda$ 
and thus  may be compared  between gauge and string theory.\foot{The 
case of  large $\J$ with fixed $ {\S \over \J}$ is familiar ``fast string'' limit.  
The two   limits --   (i)  taking $\J$ large for fixed $\S$  
 and (ii)  taking  $\J$ large for fixed   $u= {\S \over \J}$
and then expanding in small $u$ --  lead to the same result as the 
dependence on $u$ happens to be 
analytic.}
Due to the large  underlying  symmetry of the 
theory 
and the   special nature of the 
large $J$ limit 
the 
 two  expansions \rf{h} and \rf{ee}  have   the same   dependence of the spins 
and  can be universally  described  by  
\be
  \g=e= E-S-J = \sum_{n=1}^\infty   { q_n \ov J^n } 
 \ , \ \ \ \ \ \    \ \ \ \ \ \  
q_n=q_n\big({\te {S\ov J}}, \lambda\big)\,.     \la{err} \ee
In the  perturbative   gauge theory one finds from   \rf{dma},\rf{h},\rf{hi}
\be \la{wee}
&& q_1 =\lambda  a_{10} \ , \ \ \ \ \  q_2 = \lambda  a_{11} \ , \ \ \ \ 
q_3 = \lambda  a_{12} + \lambda^2   a_{20} \ , \ \ \ \ 
q_4= \lambda  a_{13} + \lambda^2   a_{21}  \ , \ \ \ \ \\ 
&&  q_5 =  \lambda  a_{14} + \lambda^2   a_{22}  +   \lambda^3   a_{30} + \lambda^4   \td a_{40} +
 \dots\ , \ \  \ \ \ ...\ \ \
\la{pty}  \ee
while in the perturbative  string theory   eqs. \rf{mai},\rf{cv},\rf{y}  give
\be \la{see}
&& q_1 =\lambda  b_{00} \ , \ \ \ \ \  q_2 = \lambda b_{10} \ , \ \ \ \ 
q_3 = \lambda^2  b_{02}  +  \lambda   b_{20}     \ , \ \ \ \ 
q_4=  \lambda^2  b_{ 12}    + \lambda   b_{30}  \ , \ \ \ \ \\ 
&&  q_5 =  \lambda^{3} b_{04} + \lambda^{5/2} b_{13} +  \lambda^{2} b_{22} + \lambda^{3/2} b_{31}+
 \lambda b_{40}+  \lambda^{1/2}  \ccm_{50} +  \dots\ ,
\ \  \ \ \ ...\ \ 
\la{ptyy}  \ee
Here $b_{13},b_{31},\ccm_{50}$, etc., are related to the 
quantum phase contributions.\foot{In particular,
 the dressing phase  corrections are responsible for 
  non-analytic terms with half-integer powers of $\lambda$ 
  and for 
  the  resolution \ci{bt,bes} of the 
  ``3-loop disagreement'' \ci{cal,ss}.}  
The functions $q_1,...,q_4$  turn out to be  protected 
 (i.e. exactly given by  
linear or quadratic  functions of $\lambda$  at   both large and small $\lambda$).\foot{
This non-renormalization of $q_1,q_2,q_3,q_4$ 
 should be due to  the underlying supersymmetry  of the  large $J$ expansion 
and a particular  structure of the  ABA  \ci{bes}.  
Equivalently,  it  may be 
considered to be   a consequence of   
exactness of the coefficients of the first 
few leading ``protected'' low-derivative
terms in the underlying effective Landau-Lifshits type action  \ci{kru,mtt1,mtt2}.}
At the same time, $q_5,q_6, ...$ are 
already 
non-trivial ``interpolating''  functions of $\lambda$. 
For example,  \rf{pty} and \rf{ptyy} represent weak-coupling and
 strong coupling expansions 
of the same   $q_5$, with dots standing for further 
infinite number of contributions coming from 
the quantum  dressing phase in ABA.

 
 The non-renormalization property of  $q_1,q_2,q_3,q_4$ implies
that the  corresponding { coefficient  functions} of  
 $S \ov J$  should be the same 
in both  the  gauge-theory and the  string-theory expansions, i.e. 
\be \la{maa}
 a_{10}=  b_{00} \ , \ \ \ \   a_{11}  = b_{10} \ , \ \ \ \ \
 a_{12} =    b_{20} \ , \ \ \ \     
 a_{20}=    b_{02} \ , \ \ \ \ \ \  a_{13} =  b_{30} \ , \ \ \ \ \    a_{21}  =     b_{12} \ . \ee 
We thus get {\it six} ``non-renormalization theorems'', 
relating low-loop  gauge theory coefficient functions  to 
low-loop string theory ones, i.e. 
six infinite  families of relations  between coefficients 
in the expansions in power series in $\sj$. 
Explicitly, as follows from \rf{maa} and \rf{ma}, \rf{cv} we 
find an  infinite number of 
relations between the coefficients in \rf{dma} and \rf{mai}   ($r=1,2,...$) 
\be
 &&a_{1 q;r} =  b_{q 0;r} \ ,\  \ \ \ \ \ \ {\rm i.e.}  \ \ \ \ \ \ 
  \ca_{1,q+r;r} =  \cm_{q r;r}   \ , \ \ \ \  \ \ \ \    q=0,1,2,3 \ , \no \\
  &&   a_{2 q;r  }   =  b_{q 2;r}\ ,\  \ \ \ \ \ \ {\rm i.e.}  \ \ \ \ \ \ 
         \ca_{2,q+r;r}=  \cm_{q, 2+r;r} \ , \ \ \ \ \  q=0,1     \la{qq} \ . \ee 
The matching of the  1-loop gauge and tree-level string coefficient functions 
 $a_{10}=  b_{00}$ was  demonstrated in \ci{bmsz,ft3,bfst}; the matching 
of the one-loop  gauge and the one-loop string coefficients 
$ a_{11}  = b_{10} $ was seen in \ci{btz}; the matching between the 1-loop 
gauge and 2-loop string coefficients  $  a_{12} =    b_{20}$   was  
checked (on the example of fast 
large-spin folded  string) in \ci{grrt}.

The relations \rf{maa}  should be universal, i.e. should not depend on a
 particular string solution (and 
 should apply  to generic states, e.g., for  $\mksu$  sector states).
Some   of these relations will be  checked below using 
explicit tree-level plus 1-loop string results and 1-loop and 2-loop gauge theory 
results for the folded string states. 
 We may then  use them to make predictions, e.g.,  
   about higher loop  string  coefficients from the
independent knowledge of gauge-theory  coefficients. 
For example,  relations originating from $a_{12}= b_{20},\  a_{13}= b_{30}$  
may   be used to get information about
some  2-loop and 3-loop  string coefficients from the knowledge 
of the  1-loop gauge theory coefficients. 

Again, starting with   $q_5$  the functions $q_n$ in \rf{err}
get non-trivial all-order dependence on $\lambda$ 
and thus their expansions at small and large $\lambda$ at fixed $ u=
{S \ov J}= { \S \ov \J }$ look different  and the 
coefficients there cannot be matched.\foot{In particular, in \rf{pty},\rf{ptyy}  
  $a_{30} \not= b_{04} $  which is an example of ``3-loop disagreement''.}

 The above discussion of the structure of $q_n$ in \rf{err} 
 applies for generic values of $\S \ov \J$, i.e. not only for 
 ${\S \ov \J}  \ll 1 $ but also for ${\S \ov \J}  \gg 1 $, 
 e.g., in the {\em fast long string} limit $\S\gg \J \gg \ln\S$ \cite{bgk,ftt}.
In this limit integer powers  of ${S \ov J}$
 in the expansion of the energy  get replaced by powers of $\ln {S \ov J}$.
 For example, the analog of $a_{12}=b_{20}$  non-renormalization relation in 
 \rf{maa} was checked in this limit in \ci{volin1,volin2}. 
 The first unprotected function $q_5$ in \rf{err}
here  has the structure \ci{ftt}
\be
q_{5} = d (\lambda)\, \ln^{6}\frac{S}{J} + ...
\,  \qquad d (\lambda) = \left\{
\begin{array}{ll}
\lambda^3(d_{0}+d_{1}\,\lambda+\dots), & \lambda \ll 1,   \\ 
\displaystyle \lambda^3(1+\frac{16}{3\,\sqrt{\lambda}}+\dots), & \lambda \gg 1 
\end{array}
\right.
\ee
where $d_{0}\not=1, \ d_{1}\sim \zeta({3})$. 

One of our aims  here 
is to understand the implications and possible 
 extensions of  the non-renormalization relations (\ref{maa}).
 A new motivation comes from the  recent observation \ci{bas} 
 of the special role  of the linear in spin terms in the energy 
 -- the corresponding coefficient (``slope'' function)
   turns out not to receive contributions 
     from  the dressing phase in ABA  \ci{kol,bas2}. That means 
   that while in general the functions $q_5, q_6, ...$ in \rf{err} are 
   non-trivially renormalised, their parts {\it linear}
    in $S\ov J$ are effectively protected, 
   i.e.  can be directly recovered  either from the gauge theory or string theory 
   perturbative expansions  without any resummation involved.
As was  originally proposed in \cite{bas} and further discussed in \cite{gva,bgmrt},
we can consider the formal
 expansion of the  string 
 energy in  small semiclassical spin parameter $\S$. 
 Expressing  $\S$  then as  ${S\ov \sqrt{\lambda}}$ we get a formal ``small $S$'' expansion 
\be 
&&E^2 = J^2 + \v_1 (\J, \sqrt{\lambda})\  S +   \v_2 (\J, \sqrt{\lambda})\  S^2 + ... \ ,
\ \ \  \ \ \  \ \ \  \ \ \ E= J + {\v_1 \ov 2 J}  S + ... \ , 
 \la{sl} \\  
&&
\v_1 = \sqrt{\lambda}   \v_{1,0}+  \v_{1,1} + {  \v_{1,2} \ov \sqrt{\lambda}} + ...\ , \ \ \ \ \ 
\v_2 =   \v_{2,0} + {  \v_{2,1} \ov \sqrt{\lambda}} + ...\ , \ \ \ \   \v_{n,k}=\v_{n,k}(\J)  \ . \la{be}
\ee
Similar relation with $h_1=h_1(J,\lambda)$  can be found on the gauge theory side by a formal 
analytic continuation to the region where  $S \ll 1$.
At weak coupling, $\lambda\ll 1$ and for  $J\gg 1$   we get 
\be 
 h_1 = 2 J  + \sum_{n=1}^\infty  { c_n(\lambda) \ov J^n } \ ,\la{hw}
 \ee
 where the   functions $c_{n}(\lambda)$ are finite polynomials in $\lambda$, e.g., 
 \be
 \label{val-cn}
 c_{1}  = \lambda, \ \ \ 
 c_{2} =  -\lambda, \ \ \ 
c_{3}   =    \lambda -\frac{\lambda ^2}{4}, \ \ \ 
  c_{4}  =   -\lambda +\lambda^2, \ \ \ 
 c_{5}   =   \lambda -\frac{11 \lambda ^2}{4}+\frac{\lambda ^3}{8}\ \ \ \ \ 
 \\   c_{6}   =    -\lambda +\frac{13 \lambda ^2}{2}-\lambda ^3,\ \  \
 c_{7}   =    \lambda -\frac{57
   \lambda ^2}{4}+5 \lambda ^3-\frac{5 \lambda ^4}{64},\ \ \ 
c_{8}   =    -\lambda +30 \lambda
   ^2-\frac{81 \lambda ^3}{4}+\lambda ^4, \ \  ... \no  
   \ea
  At strong coupling or semiclassical 
  string theory, with  $ \sqrt{\lambda} \gg 1$   and  $ \J= { J \ov \sqrt{\lambda}}  \gg 1$, we have 
 \be 
 h_1 =2 \sqrt{\lambda} \J  + \sum_{n=1}^\infty  { a_n(\sqrt{\lambda}) \ov \J^n }
 =      2 J  + \sum_{n=1}^\infty  { \tilde c_n(\lambda) \ov J^n }\ .\la{hs}
 \ee
Due to the  absence of the dressing phase contribution to $h_1$ 
 \ci{bas,kol,bas2}
 one may   expect  that $\tilde c_n(\lambda)$  should 
  also given by  same {\it finite} polynomials 
 in $\lambda$, without any   resummation, 
\be
\label{conj}
c_n(\lambda)=  \tilde c_n(\lambda)\  .
\ee
This  provides a  non-trivial extension 
of the ``non-renormalization'' relations in \rf{wee},\rf{see}.
This direct relation can indeed be proved starting from 
the explicit expression of the 
slope $h_1$ \cite{bas}
valid for all $\lambda$   and $J$ (below $I_{k}(x)$ is the modified Bessel function of the first kind)
\be
\label{h1}
h_{1} (J,\lambda)= 2\,\sqrt{\lambda}\,\frac{d}{d\sqrt\lambda}\ln I_{J}(\sqrt{\lambda}) 
=  2J + 2 \sqrt{\lambda}\, { I_{J+1} (\sqrt{\lambda}) \ov I_J (\sqrt{\lambda})}   \ . 
\ee
It obeys the differential equation
\be
\label{h1-diffeq}
\frac{dh_{1}}{d\lambda}+\frac{1}{4\lambda} h_{1}^{2}-\frac{J^{2}}{\lambda}-1=0.
\ee
If we replace $h_{1}$ here 
by its expansion (\ref{hw}),  we immediately determine 
 the functions $c_{n}(\lambda)$  in (\ref{val-cn}).
They are polynomials  valid 
for  all  values of $\lambda$ since they are derived from  (\ref{h1-diffeq}) 
which is exact in $\lambda$.~\footnote{
From (\ref{h1-diffeq}), we deduce the 
following recursion relation for the polynomials $c_{n}(\lambda)$
\be\no
c_{1} = \lambda, \qquad c_{n} =
-\lambda\,c_{n-1}'-\frac{1}{4}\,\sum_{m=1}^{n-2}\,c_{m}\,c_{n-m-1} \ .
\ee
}

In the case of higher order functions $h_{k}$ with $k>1$ in \rf{sl} 
the expansion like \rf{hw} will  have non-trivial 
coefficients starting with $c_5$: they will correspond to  $S^2, S^3,...$ 
terms in  $q_n$ in \rf{err}   and thus are expected to be 
given by interpolating functions having different form when expanded  
at weak and at strong coupling. 


Similar  considerations apply to  strings moving in $S^5$, e.g., for
 folded string in the  $\mksu$ sector
 having spin $J'$ and orbital momentum $J$: the analog of \rf{sl} is then 
\be 
E^2 = J^2 + \u_1 (\J, \sqrt{\lambda})\  J' +   \u_2 (\J, \sqrt{\lambda})\  J'^2 + ... \ ,
\ \ \  \ \ \ \ \  E= J + {\u_1 \ov 2 J}  J' + ... \ .
 \la{su} 
\ee 
There is a simple  observation that 
allows one to determine the $\mksu$ sector slope  $\u_1 $ in terms of the 
$\mksl$ sector one  $h_1$  in \rf{sl},\rf{h1}.
Using that  the two  folded string  solutions  are  related   by 
an analytic  continuation   \ci{bfst},  
 it is possible to derive  the following relation 
 between the two slope functions:\foot{Here we formally 
 use the same notation for the functions  of $(\J,\sqrt{\lambda})$ and of $(J,\sqrt{\lambda})$.}
\be 
\u_1(\J, \sqrt{\lambda}) = - \v_1 (\J, -\sqrt{\lambda}) \ , \ \ \ \ \ {\rm i.e.}  \ \ \ \ \
\u_1(J, \sqrt{\lambda}) = - \v_1 (-J, -\sqrt{\lambda}) \ ,\ \ \ \    J= \sqrt{\lambda} \J   \ .
\la{rr}  \ee
Indeed, given a 
 classical  solution  for a string moving  in  $AdS_3\times S^1  $  
  with energy and spins   $(\E, \S;  \J)$
  it can be  related (by an analytic continuation converting   $AdS_3  $   into $S^3$)
  to  a    classical solution in $R \times S^3$   with  
  the energy and spins $(\td  \E; \J', \td \J)$
such that $ \E = - \J,  \   \td \J= -\E,\  \td J_1\equiv  \J'= S $. 
Since this  continuation involves  setting the radial direction 
$\rho$  in  $AdS_3$   equal to $i \theta$  where $\theta$  is an angle
 in $S^3$, 
 the action changes sign. This sign change 
 can  be compensated 
by  reversing  the sign of the string  tension \ci{rtt}, $\sqrt{\lambda} \to -\sqrt{\lambda}$, 
 thus ensuring 
  that the quantum corrections  to the two solutions are also in correspondence. 
As a result,  the relations   between the parameters  of the $\mksu$  and $\mksl$   solutions are 
\be \td  \E = - \J, \ \ \ \ \ \   \td\J= -\E,\  \ \ \ \ \ \  \J'= \S \ , \ \ \ \ \ \ \ 
\widetilde {\sqrt{\lambda}} =-\sqrt{\lambda} 
 \ . \la{sk} \ee
Expanding the two energies in respective 
small  spins $\S$ and $\J'$  we  then get (cf. \rf{sl},\rf{su})
\be 
E^2 = J^2 + \v_1(\J, \sqrt{\lambda}) \ S + ...\ , \ \ \ \ \ \ \ \ 
\td E^2 =\td  J^2 + \u_1(\td \J, \widetilde{ \sqrt{\lambda}}) \ J'  + ...\ , \ee
which implies \rf{rr}   after using \rf{sk}.\foot{A  similar proposal for the   $\mksu$  slope,
based on $\sqrt{\lambda} \to - \sqrt{\lambda}$, $J \to - J$  in the $\mksl$  slope 
  was independently
   made in  \ci{kol} by starting with the ABA at weak coupling.} 
Below  we shall study  the 
consequences  of  \rf{rr} in detail, 
determining the  exact expression  for the  $\mksu$  slope $\u_1$
in terms of Besssel $K$-function  and its explicit  
behaviour  at weak and at strong coupling.


The rest of this paper is organized as  follows.
In section 2 we shall  check  the ``non-renormalization'' relations 
\rf{maa} listing the values of $a_{nk}$   coefficients in \rf{wee} expanded in small 
spin limit and comparing them to string theory data in Appendices A and B. 
In  section 3   we shall discuss our proposal for the slope function for the $(J',J)$
folded string state  in $\mksu$ sector.
Details of  string-theory and gauge-theory computations are summarized in Appendices A--E.

\renewcommand{\theequation}{2.\arabic{equation}}
 \setcounter{equation}{0}


\section{Check of the ``non-renormalization'' relations}

The  relations \rf{maa}  between first few  leading coefficients 
on the string and gauge theory sides  can be demonstrated 
 explicitly in small spin (or small $u$ \rf{uuu}) expansion for the folded
string in $AdS_5$ or $S^5$. Below we list  the results that follow from 
the perturbative data given in Appendices. 

\subsection{Folded string in $\mksl$  sector}

The gauge-theory expressions for the functions $a_{nk}$ in \rf{wee} 
entering 
the non-renormalization relations (\ref{maa})
 for  the $(S,J)$  folded string state in $\mksl$  sector
can be read from the results of Appendix 
 \ref{app:gauge-sl2}  (here $u=S/J$)
\ba
a_{10} &=& \frac{u}{2}-\frac{u^2}{4}+\frac{3 u^3}{16}-\frac{21 u^4}{128}+\frac{159
u^5}{1024}+\dots,  \no
\\
a_{11} &=& -\frac{u}{2}+\Big(\frac{1}{8}-\frac{\pi ^2}{12}\Big) u^2+\Big(\frac{3}{64}
+\frac{\pi ^2}{24}+\frac{\pi ^4}{180}\Big) u^3
+\Big(-\frac{99}{512}+\frac{\pi ^2}{384}-\frac{\pi ^4}{240}-\frac{\pi ^6}{1512}\Big)
u^4+\dots,\no
 \\
a_{12} &=& \frac{u}{2}+\Big(-\frac{3}{16}+\frac{\pi ^2}{4}
-\frac{\pi ^4}{90}\Big) u^2+\Big(\frac{3}{16}-\frac{43 \pi ^2}{192}
-\frac{\pi ^4}{120}+\frac{11 \pi ^6}{3780}\Big) u^3+\dots, \la{1}\\
a_{13} &=& -\frac{u}{2}+\Big(\frac{5}{32}-\frac{19 \pi ^2}{48}+
\frac{2 \pi ^4}{45}-\frac{\pi ^6}{315}\Big) u^2+\dots, \no \\
a_{20} &=& -\frac{u}{8}-\frac{u^2}{4}+\frac{11 u^3}{32}-\frac{27 u^4}{64}+\frac{1041
u^5}{2048}+\dots,\no
 \\
a_{21} &=& \frac{u}{2}+\Big(\frac{5}{8}+\frac{\pi ^2}{24}\Big) u^2+
\Big(-\frac{11}{16}+\frac{\pi ^2}{12}\Big) u^3
+\Big(\frac{561}{1024}-\frac{29 \pi ^2}{256}-\frac{17 
\pi ^4}{2880}-\frac{\pi ^6}{3024}\Big) u^4+\dots  \ . \no 
\ea
These expressions  
 can be compared with the available information  about 
 string theory coefficients $b_{nk}$ in \rf{see} 
  summarized in Appendix 
\ref{app:string-sl2}.  The relations  (\ref{maa}) indeed  hold in all cases 
where there is 
string data for  comparison  to be   made. 
In addition, the above gauge-theory functions contain also
 terms that 
can be directly tested at the moment and thus provide  
  2- and 3-loop string predictions: such are  
  the terms in $a_{12} = b_{20}$  and $a_{13}=b_{30}$.

In Appendix \ref{app:gauge-sl2}, 
we also compute the leading dressing phase correction 
to the gauge-theory anomalous dimension \rf{k}. The first term in 
 (\ref{lo-dressing}) is proportional to 
$\frac{\lambda^{4}}{J^{5}}\,
(\frac{S}{J})^{2}$ and thus  contributes to $q_{5}$ in (\ref{pty}).
 This indicates   that in contrast to $q_1,q_2,q_3,q_4$ functions, starting with 
 order $S^2$ terms  (contributing to higher slope $h_2, ...$ functions in \rf{sl}) 
  the function  $q_{5}$ is not given  by a finite polynomial 
  as would  be the case if one were to  use the BDS ansatz. 
  

\subsection{Folded string in  $\mksu$   sector}

The functions $a_{nk}$ in \rf{wee} in the 
case of the $(J',J)$  folded string in $\mksu$ sector 
are computed in Appendix  \ref{app:gauge-su2}  (here $u=J'/J$)
\ba
a_{10} &=& \frac{u}{2}-\frac{3 u^2}{4}+\frac{15 u^3}{16}-\frac{139 u^4}{128}+\frac{1239
 u^5}{1024}+\dots,\no
  \\
a_{11} &=& \frac{u}{2}-\Big(\frac{11}{8}+\frac{\pi ^2}{12}\Big) u^2+
\Big(\frac{157}{64}+\frac{7 \pi ^2}{24}-\frac{\pi ^4}{180}\Big) u^3
- \Big(\frac{1899}{512}+\frac{239 \pi ^2}{384}-\frac{17 \pi ^4}{720}+
\frac{\pi ^6}{1512}\Big) u^4+\dots \no \\
a_{12} &=& \frac{u}{2}+\Big(-\frac{29}{16}-\frac{\pi ^2}{4}+
\frac{\pi ^4}{90}\Big) u^2+\Big(\frac{17}{4}+\frac{197 \pi ^2}{192}
-\frac{23 \pi ^4}{360}+\frac{11 \pi ^6}{3780}\Big) u^3+\dots,\la{2}  \\
a_{13} &=& \frac{u}{2}+\Big(-\frac{75}{32}-\frac{19 \pi ^2}{48}+
\frac{2 \pi ^4}{45}-\frac{\pi ^6}{315}\Big) u^2+\dots, \no\\
a_{20} &=& -\frac{u}{8}+\frac{u^2}{4}-\frac{9 u^3}{32}+\frac{7 u^4}{64}+\frac{761
u^5}{2048}+\dots, \no
\\
a_{21} &=& -\frac{u}{2}+\Big(\frac{15}{8}+\frac{\pi ^2}{24}\Big) u^2
+\Big(-\frac{67}{16}-\frac{\pi ^2}{12}\Big) u^3+\Big(\frac{7449}{1024}-\frac{29 \pi ^2}{256}+\frac{13 \pi
 ^4}{960}-\frac{\pi ^6}{3024}\Big) u^4+\dots\ . \no
\ea
Again, they can be compared with the  available 
string theory data  for the $b_{nk}$ in \rf{see}
given in Appendix \ref{app:string-su2} and 
the relations  (\ref{maa})  hold in all cases.

\renewcommand{\theequation}{3.\arabic{equation}}
 \setcounter{equation}{0}
 
\section{A proposal for the slope function  in the $\mksu$ sector}

Let us now study the 
proposal for slope function $\td h_1$  in the $\mksu$ sector
implied by the relation \rf{rr} to the slope $h_1$ in the  $\mksl$ sector.
The relation \rf{rr} was motivated from strong coupling so 
the precise definition of $\td h_1$ at weak coupling 
may  need extra input.  

Starting  with 
the strong-coupling expansion of $h_{1}$  in \rf{h1} for fixed $\J$ 
\ba
&& h_{1}(\mc J, \sqrt\lambda) =2 \sqrt{\mathcal{J}^2+1}\,\sqrt\lambda-\frac{1}{\mathcal{J}^2+1}+
\frac{4 \mathcal{J}^2-1}{4 \big(\mathcal{J}^2+1\big)^{5/2}}\,\frac{1}{\sqrt\lambda} \no\\  && \ \ \ 
-
\frac{4 \mathcal{J}^4-10 \mathcal{J}^2+1}{4 \big(\mathcal{J}^2+1\big)^4}\,\frac{1}{(\sqrt
\lambda)^{2}}
 + \frac{8 \mathcal{J}^2 \big(8 \mathcal{J}^4-70 \mathcal{J}^2+57\big)-25}{64 \big(
\mathcal{J}^2+1\big)^{11/2}}\,\frac{1}{(\sqrt\lambda)^{3}} \no \\
&& \ \ \ -\frac{16 \mathcal{J}^8-368 \mathcal{J}^6+924 \mathcal{J}^4-374 \mathcal{J}^2+13}{16 
\big(\mathcal{J}^2+1\big)^7}\frac{1}{(\sqrt\lambda)^{4}}+\dots,\la{4}
\ea
eq. \rf{rr} implies that 
to get the corresponding  expansion 
of $\td h_{1}$  one is to change $\sqrt{\lambda}\to -\sqrt{\lambda}$  and change the overall sign, i.e. 
\ba
&&\td h_{1}(\mc J, \sqrt\lambda) = 2 \sqrt{\mathcal{J}^2+1}\,\sqrt\lambda+\frac{1}{\mathcal{J}^2+1}+
\frac{4 \mathcal{J}^2-1}{4 \big(\mathcal{J}^2+1\big)^{5/2}}\,\frac{1}{\sqrt\lambda}\no \\
&& \ \ \ +
\frac{4 \mathcal{J}^4-10 \mathcal{J}^2+1}{4 \big(\mathcal{J}^2+1\big)^4}\,\frac{1}{(
\sqrt\lambda)^{2}}+ \frac{8 \mathcal{J}^2 \big(8 \mathcal{J}^4-70 \mathcal{J}^2+57\big)-25}{64 
\big(\mathcal{J}^2+1\big)^{11/2}}\,\frac{1}{(\sqrt\lambda)^{3}}\nonumber \\
&& \ \ \  +\frac{16 \mathcal{J}^8-368 \mathcal{J}^6+924 \mathcal{J}^4-
374 \mathcal{J}^2+13}{16 \big(\mathcal{J}^2+1\big)^7}\frac{1}{(\sqrt\lambda)^{4}}+\dots. \la{km}
\ea
Like $h_1$ in \rf{h1},\rf{4}, the  $\mksu$ slope $\td h_{1}(J, \lambda) $ admits also 
a regular expansion at large $\sqrt{\lambda}$ and fixed $J$, that follows also from \rf{km}
upon setting $\J= {J \ov \sqrt{\lambda}}$ and re-expanding in $1 \ov \sqrt{\lambda}$ 
\be 
\la{sh}
\td h_{1}(J, \sqrt\lambda)= 
2\sqrt\lambda + 1-  \frac{1}{\sqrt\lambda}\big({1 \ov 4} - J^2  \big)+\frac{1}{(\sqrt{\lambda})^2}\,
\big({1 \ov 4}-J^{2}\big) + ...   \ , \ee
which is in agreement with expectations in \ci{bgmrt}. 
Since the  expansion \rf{km},\rf{sh} depends on {\it even} powers of $J$ only, the
relation \rf{rr} implies that 
it is  the same as the one for the $\mksl$ slope  $h_1$  in \ci{bas}
but  with the signs of the 
terms with even powers of $1 \ov \sqrt{\lambda}$ reversed.

One can compare the  three-loop gauge theory  data given in Appendix C 
for the $\mksl$ slope function $h_1$ 
 to show that it  agrees with the exact expression for  the 
coefficients $c_{1},...,c_8$ given explicitly in  (\ref{val-cn}). This is a consequence of the 
fact that $h_1$ does not receive  contributions from the 
dressing phase. Inspecting  similar three-loop data  for the coefficients in 
the slope $\td h_{1}$  in the $\mksu$ sector \rf{su}
collected in Appendix \ref{app:gauge-su2}, we find  that they are in agreement
 with the proposed relation (\ref{rr}).

Using  the explicit 3-loop  gauge theory data 
the slopes 
$h_{1}$ and $\td h_{1}$ can be resummed to all orders in $1/J$ and take the form\footnote{
The expression  (\ref{eq:sl2-weak-h1}) is  of course also obtained from the 
exact expression in (\ref{h1}).}
\ba
\label{eq:sl2-weak-h1}
h_{1}(J, \lambda) = 
2\,J+\frac{\lambda}{J+1}-\frac{\lambda^{2}}{4\,(J+1)^{2}\,(J+2)} 
 +\frac{\lambda^{3}}{8\,(J+1)^{3}\,(J+2)\,(J+3)}+\cdots, \\
\label{wh1}
\td h_{1}(J, \lambda) = 
2\,J+\frac{\lambda}{J-1}-\frac{\lambda^{2}}{4\,(J-1)^{2}\,(J-2)} 
 +\frac{\lambda^{3}}{8\,(J-1)^{3}\,(J-2)\,(J-3)}+\cdots.
\ea
We  observe that these  two expressions 
 are indeed related by (\ref{rr}), i.e.
 by $h_{1}(J, \lambda) = - \td h_{1}(-J, \lambda)$
 as functions of integer powers  of $\lambda$. 
 The expression   (\ref{wh1})
 is  the same as  found  in \cite{kol}.

Let us now address the question about an exact expression for 
 $\td h_{1}(J, \lambda) $ that 
   correctly  interpolates  between the correct 
    weak-coupling  and strong-coupling
expansion. This question turns out to be non-trivial: 
\begin{enumerate}
\item[(a)] At strong coupling,  the transformation (\ref{rr}) cannot  be directly 
implemented by 
doing the replacement $\sqrt{\lambda}\to -\sqrt{\lambda}$ in the exact expression 
(\ref{h1}).
\footnote{A simple example that explains 
why this  is not so is the following large $x$ expansion
\be
\frac{x^{2}\pm \sqrt{x^{2}+1}}{\sqrt{x^{2}+2}}=
x\pm 1-\frac{1}{x}\mp\frac{1}{2 x^2}+\frac{3}{2 x^3}
\pm\frac{7}{8 x^4}-\frac{5}{2 x^5}\mp\frac{25 }{16 x^6}
\cdots. \no
\ee
It shows that the transformation that changes half of the 
series has nothing to do with sign flip of  $x$. This is due 
to the branch point at infinity and to the fake odd powers of 
$x$ arising from square roots. 
 }
\item[(b)] At weak coupling, the expansion (\ref{wh1})  breaks down at some order
in $\lambda$  for any positive integer $J$.
\end{enumerate}

To try to resolve  these  problems, we 
recall that 
 the function
$Y_{J}(x) = I_{J}'(x)/I_{J}(x)$  (here prime is derivative over  $x=\sqrt\lambda$),
entering the expression \rf{h1}  for the exact $\mathfrak{sl}(2)$ slope \ci{bas}
 obeys the relations 
\be
Y_{J}' = 1+\frac{J^{2}}{x^{2}}-\frac{Y_{J}}{x}-Y_{J}^{2},\qquad  \qquad Y_{J}(+\infty)=1\ .\la{yy}
\ee
Changing sign of $x$ as instructed  by \rf{rr}, 
 we are interested in the solution $Z_{J}(x)$ to the following conditions
\be
-Z_{J}' = 1+\frac{J^{2}}{x^{2}}+\frac{Z_{J}}{x}-Z_{J}^{2},\qquad  \qquad Z_{J}(+\infty)=1 .\la{zz}
\ee
Here we used that  both  $ h_1$ and $\td h_1$, 
\be h_1 =2 \sqrt{\lambda}  Y'_{J}\ , \ \ \ \ \ \ \ \ \ \ \ \ \
\td h_1 =2 \sqrt{\lambda}  Z'_{J} \ ,  \ee  should have the strong-coupling asymptotics 
$2 \sqrt{\lambda} $ at  fixed $J$ (see \rf{4},\rf{km}  where in this limit $\J=0$).
This implies the  boundary condition $Z_{J}(+\infty)=1$. 
The unique 
solution of \rf{zz} is given in terms of  the modified Bessel function 
of the second kind $K_{J}$, i.e.  
$Z_{J}(x) =(\ln K_{J}(x) )'$. Thus our proposal for 
$\td h_{1}(J, \lambda)$ is\foot{An equivalent form 
of this expression is  $
\td h   = -2\,\sqrt{\lambda}\,\frac{d}{d\sqrt{\lambda}}\ln H^{(1)}_{J}(i\sqrt{\lambda}) $
where $H_{J}^{(1)}$ is the first Hankel function. 
This follows from the relation 
$K_J(x) = {\pi \ov 2}   i^{J+1}  H^{(1)}_{J}(i x)$.
We thank B. Basso for a suggestion to express $\td h_{1}$ in terms of $K_J$.
Note  that 
$K_{J}(x) = \frac{\pi}{2\sin(\pi J)}\Big[ {I_{-J}(x)-I_{J}(x)}\Big]$
and that 
$h_1$   in \rf{h1} may be written also as \ci{bas}\ \ 
$h_1= - 2J + 2\sqrt{\lambda}\, { I_{J-1}(\sqrt{\lambda})\ov I_{J}(\sqrt{\lambda})} $.
}
\be
\td h_{1}(J, \lambda)  &=& -2\,\sqrt{\lambda}\,\frac{d}{d\sqrt{\lambda}}\ln K_{J}(\sqrt{\lambda}) \no \\
&=&  2J + 2\sqrt{\lambda}\, { K_{J-1}(\sqrt{\lambda})\ov K_{J}(\sqrt{\lambda})}  \ . \la{hh}
\ee
One can  check immediately that the strong coupling expansion 
of this function at fixed $\J$ 
is in agreement  with (\ref{km}),  solving  the above  problem (a).

 As an illustration, to  
  compare the expressions for the  $\mksl$ \rf{h1} and the 
  proposed $\mksu$ \rf{hh} 
  slope functions 
 we plotted them  together for $J=3$  in Figure 1.

\begin{figure}[H]
\begin{center}
\includegraphics[scale=0.8]{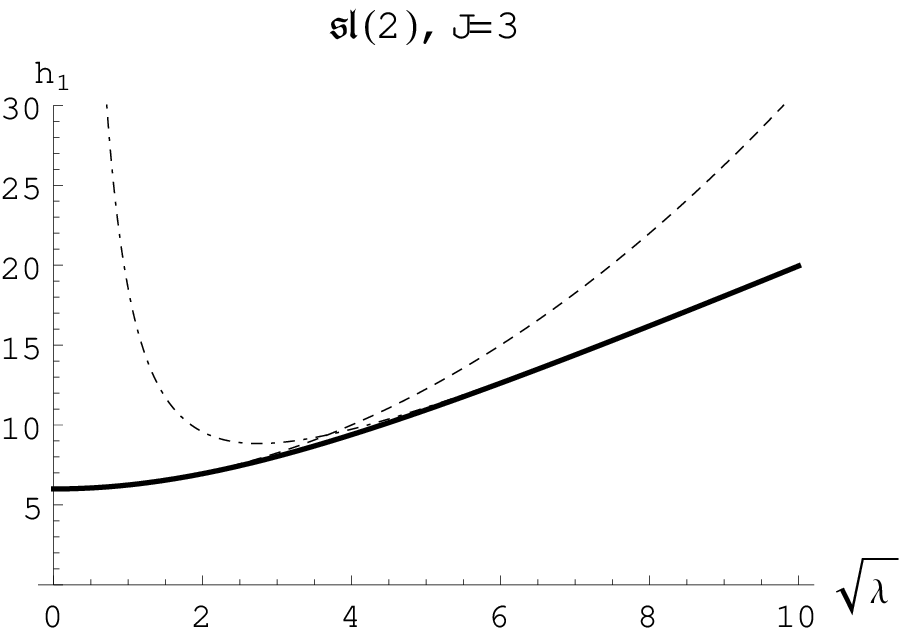}
\includegraphics[scale=0.8]{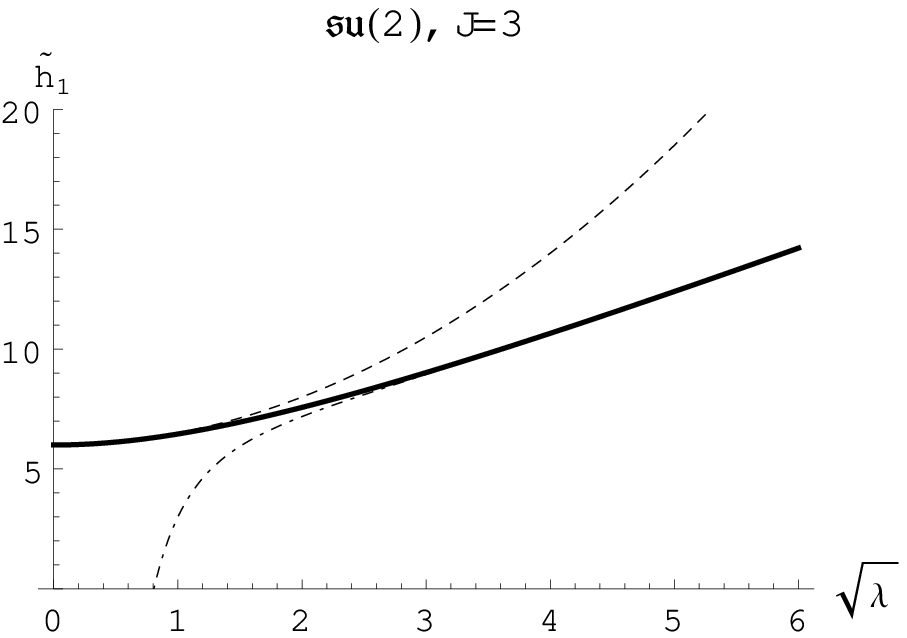}

\

\vskip 4pt

\includegraphics[scale=1.2]{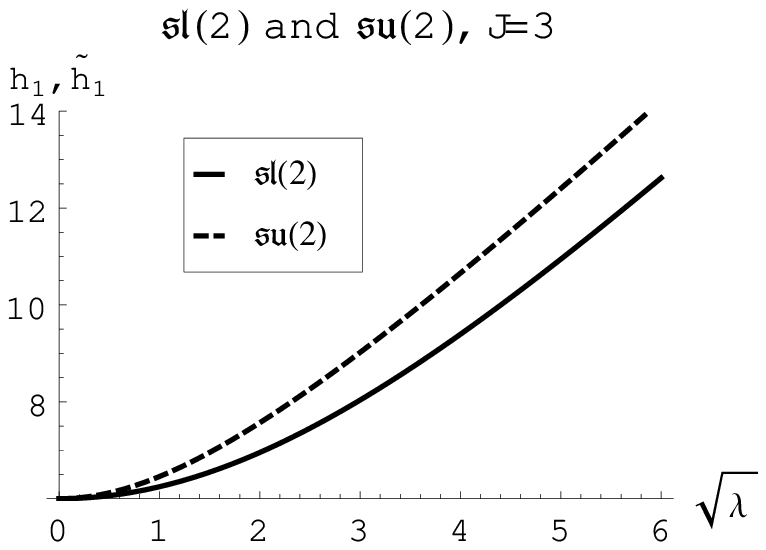}

\caption{Slope  functions in $\mksl$ and $\mksu$ sectors at $J=3$}
\label{fig:slope}
\end{center}
\end{figure}


Here on the upper-left plot 
  the solid line represents  the  $\mksl$  exact  slope function \rf{h1}, 
 the dashed line is the one-loop 
 weak coupling asymptotics $2\,J + \frac{\lambda}{J+1}$ and the  
dot-dashed line is the four-term strong-coupling asymptotics 
$2\sqrt\lambda-1- \frac{1}{\sqrt\lambda}\,({1 \ov 4}-J^{2})-\frac{1}{(\sqrt{\lambda})^2}
\,({1 \ov 4} -J^{2})$ (cf. \rf{4}). 
On the  upper-right plot 
 the solid line is our proposed expression for the $\mksu$ slope function 
 \rf{hh},  the  dashed line is the one-loop weak coupling expansion 
$2\,J+\frac{\lambda}{J-1}$  and the 
dot-dashed line is the four-term strong-coupling expansion 
$2\sqrt\lambda+1- \frac{1}{\sqrt\lambda}\,({1 \ov 4}-J^{2} )+\frac{1}{(\sqrt{\lambda})^2}\,
({1 \ov 4}-J^{2})$ (cf. \rf{km},\rf{sh}). 
The lower plot contains both slope curves  at the same time: 
the $\mathfrak{sl}(2)$  one (solid lower line) 
and the $\mathfrak{su}(2)$ one (dashed upper curve).

The function $\td h_1$  defined by \rf{hh}  is a smooth  function of $\lambda$ at either large
or small $\lambda$ 
for all  $J$, including integer ones (as illustrated by Figure 1 for $J=3$). 
What happens for  positive integer $J$ is that the small  $\lambda$ 
expansion \rf{wh1}  becomes asymptotic   rather than having finite 
radius of convergence (as expected for a sum of planar graphs). 
 Indeed, expanding  the function $\td h_{1}$ in \rf{hh} 
at  weak coupling
at fixed integer values of $J$ to find\foot{Here higher-order terms contain 
higher powers of $\ln \lambda$, e.g., for  $J=1$ one gets 
\be\no 
\tilde h_{1}(1, \lambda)  = 2-\lambda\,\Big(\ln\frac{\lambda}{4}+2\,\gamma_{\rm E}\Big)
+\frac{1}{4} \lambda^{2}\,\Big[\Big(
\ln\frac{\lambda}{4}+2\,\gamma_{\rm E}
\Big)\,\Big(
\ln\frac{\lambda}{4}+2\,\gamma_{\rm E}-2
\Big)+ {2}\Big]+\dots
\ee
}
\be\la{log}
\td h_{1}= \left\{
\begin{array}{lr}
\displaystyle 4+\lambda +\frac{1}{4} \lambda ^2
\big(\ln \frac{\lambda }{4}+2 \gamma_{\rm E} \big) 
+
\dots, & J=2 \\ \\
\displaystyle 6+\frac{\lambda }{2}-\frac{\lambda ^2}{16}-\frac{\lambda ^3}{128}
\big(2 \ln\frac{\lambda }{4}+4 \gamma_{\rm E} -1\big) +\dots, & J=3 \\ \\
\displaystyle 
8+\frac{\lambda }{3}-\frac{\lambda ^2}{72}+\frac{\lambda ^3}{432}+
\frac{\lambda ^4}{20736} \big(9 \ln\frac{\lambda }{4}+18 \
\gamma_{\rm E} -8\big)
 +\dots, & J=4 \\ \\
\displaystyle 
10+\frac{\lambda }{4}-\frac{\lambda ^2}{192}+\frac{\lambda ^3}{3072}-\
\frac{7 \lambda ^4}{147456}-\frac{ \lambda^5}{2359296}
\big(16 \ln \frac{\lambda }{4}+32 \gamma_{\rm E}
 -19\big)
+\dots,  & J=5 \\
\end{array}
\right.
\ee
While the  regular $\lambda^n$ terms
here  are in full agreement with (\ref{wh1}),  the 
appearance of non-analytic $\lambda^{J} \ln \lambda $  terms 
is related to problem (b), i.e. a  breakdown of the expansion  (\ref{wh1})
at positive integer $J$. 
The same  problem appears of course in the  $\mksl$ slope function $h_1$ 
in \rf{h1},\rf{eq:sl2-weak-h1}  continued to  {\it negative} integer $J$. 

In general, anomalous dimensions are functions in 
multiparameter space of $\lambda$, $J$, spins $S$ or $J'$, etc., 
and their general  behaviour  is just beginning to be understood. 
While for generic  values of the parameters  one may expect that
the gauge-theory  dimension  given  
by  a sum  of planar diagrams  should have a finite radius of convergence in $\lambda$
this expectation may break down in certain limits 
of the parameters (like in few  known  cases of IR divergences,  elimination of which 
requires a resummation of direct  perturbation theory in $\lambda$   leading to non-analytic terms in
$\lambda$, see  below). 

Indeed, one   possible reason for the appearence of the above $\ln \lambda$ terms 
 is  that the definition  of the slope function
in either $\mksl$  or  $\mksu$  sector 
at finite $J$ is non-trivial  in the first place, as it  is based 
 on a formal  analytic continuation to
small  values  of spin  $S$ or $J'$  from their  standard integer values.  
The case of $\mksu$ sector is even more subtle 
 since here the spin  $J'$ is bounded  from above 
 by the  fixed length $L= J' + J$ of the spin chain, implying potential 
  problems with  an   analytic continuation to small $J'$. 
 It is possible that  the   continuation of the $\mksu$ sector 
 anomalous dimension  to small $J'$, i.e.
 the  slope $\td h_1$  is  defined only 
 in the large $J$ limit when the bound on $J'$ becomes irrelevant
 (similar remark appeared in \ci{kol}).\foot{This  will  be consistent
  with the relation \rf{rr} 
 valid at strong coupling as  in the semiclassical expansion with  fixed $\J$
 at $\sqrt{\lambda} \ll 1$  the orbital momentum $J$ is always large.} 
 In that case  \rf{wh1},\rf{hh}  may be viewed  as  a compact way of encoding the large $J$ 
 expansion. 
 
 It may happen though  that in contrast to the $\mksl$  slope,
  the $\mksu$ slope may actually 
 receive 
 wrapping  contributions which also start,  in  general, 
  at the $(\lambda^L)_{J'\to 0} \sim  \lambda^J$ order \ci{bds}.
 Taking  them into account (by using   a TBA generalization of ABA)
  may lead to a modification of \rf{hh} that will 
  make the expansion \rf{wh1} well-defined, i.e.  
   cancel  the $\lambda^{J} \ln \lambda $  terms in \rf{log}.
   This may still be  consistent with the strong-coupling relation \rf{rr}
    between   the $\mksl$ and $\mksu$ slopes   as the wrapping  contributions may 
   turn out to be  suppressed at strong  coupling.
  
 An alternative (and more likely) 
 possibility already alluded to above 
  is 
  that the $\lambda^{J} \ln \lambda $  terms in \rf{log}  actually have a physical
   meaning being analogous to  $\lambda^n \ln \lambda +...$ 
   terms  appearing in  IR-resummed perturbation  theory (see, e.g., 
     \ci{pin,cor2}).\foot{We are grateful to B. Basso for this suggestion.} 
 Indeed, there is a  similarity  between  the expansion of \rf{hh} in \rf{log} 
 and the ladder-diagram resummed expression for the 
 $q$-$\bar q$ potential in \ci{cor2}. A formal reason for this  
  may be related to an 
  analogy \ci{cor1} between 
  the expectation value  of the cusp Wilson  loop 
   at small euclidean  angle ($\phi\to 0$) 
   and the $\mksl$ slope function $h_1$ \rf{h1} at $J=1$. 
   While the 
   $q$-$\bar q$ potential is related to a different 
    ($\phi \to \pi$ or antiparallel  lines) 
   limit of the cusp Wilson loop  \ci{df,cor2},  the  relation between 
   the expressions in the  $\phi \to 0$  and  $\phi \to \pi$ limits 
    may   be similar  to the  transformation \rf{rr}
   relating  the $\mksl$   slope $h_1$ to the $\mksu$ 
   slope $\td h_1$. One may speculate that   given that 
    the  cusp Wilson loop is described,  for generic values 
   of $\lambda$  and  $\phi$,  by an integrable TBA system \ci{cor3,bad}, it  
   may    admit an  exact representation  in terms of Bessel functions  
   not only for $\p \to 0$  \ci{cor1,cor3}   but also for $\p \to \pi$.

\subsection*{Acknowledgments }

We would like to thank B. Basso, N. Beisert, N. Gromov, G. Macorini,  R. Roiban, 
and D. Volin
for  useful discussions. 
The work of AAT was supported by the ERC Advanced  grant No.290456. 
 
 While  this paper  was  in preparation  a similar proposal for 
 the $\mathfrak{su}(2)$ slope 
  was made in \ci{kol}.

\appendix
\numberwithin{equation}{section}

\section{String theory data for the $\mksl$ folded string}
\label{app:string-sl2}

\subsection{Classical charges and large $\J$ limit}

The classical charges of the $\mathfrak{sl}(2)$ folded string \ci{gkp,ft1}
can be written in parametric form 
as the following combination 
of elliptic integrals depending on the cut endpoints $a, b$ and 
read \cite{bfst}
\ba
\label{sl2-charges}
\mathcal S &=& \frac{1}{2\,\pi}\frac{ab+1}{ab}\,\Big[b\,\mathbb E\Big(1-\frac{a^{2}}{b^{2}}\Big)
-a\,\mathbb K\Big(1-\frac{a^{2}}{b^{2}}\Big)\Big], \nonumber \\
\mathcal J &=& \frac{1}{\pi}\,\frac{1}{b}\,\sqrt{(a^{2}-1)(b^{2}-1)}\,\mathbb K\Big(1-\frac{a^{2}}{b^{2}}\Big). \\
\E_{0} &=& \frac{1}{2\,\pi}\frac{ab-1}{ab}\,\Big[b\,\mathbb E\Big(1-\frac{a^{2}}{b^{2}}\Big)
+a\,\mathbb K\Big(1-\frac{a^{2}}{b^{2}}\Big)\Big],\nonumber
\ea
where $\E_{0} = \J+\S+e_{0}(\S, \J)$ 
as in  (\ref{stri}).
The small $\mc S$ expansion of $\E_{0}$ at fixed $\mc J$ reads
 \ba
 \label{eq:sl2-classical}
&& \E_{0}= \mathcal J+\frac{\sqrt{\mathcal J^{2}+1}}{\mathcal J}\,\mathcal S-\frac{\mathcal J^{2}+2}{4\,\mathcal J^{3}(\mathcal J^{2}+1)}\,\mathcal S^{2}+ 
 \frac{3\,\mathcal J^{6}+13\,\mathcal J^{4}+20\,
 \mathcal J^{2}+8}{16\,\mathcal J^{5}\,(\mathcal J^{2}+1)^{5/2}}\,\mathcal S^{3}+\dots
\ea
Taking the large $\J$ limit of (\ref{eq:sl2-classical}), we obtain
\ba
\label{eq:E0}
\E_{0} &=& 
\mathcal{J}+\mathcal{S}+\frac{\mathcal{S}}{2 \mathcal{J}^2}-\frac{\mathcal{S}^2}{4 \mathcal{J}^3}+\frac{\frac{3
   \mathcal{S}^3}{16}-\frac{\mathcal{S}}{8}}{\mathcal{J}^4}+\frac{-\frac{21
   \mathcal{S}^4}{128}-\frac{\mathcal{S}^2}{4}}{\mathcal{J}^5}+\frac{\frac{159 \mathcal{S}^5}{1024}+\frac{11
   \mathcal{S}^3}{32}+\frac{\mathcal{S}}{16}}{\mathcal{J}^6} \nonumber\\
   &&+\frac{-\frac{315 \mathcal{S}^6}{2048}-\frac{27
   \mathcal{S}^4}{64}+\frac{\mathcal{S}^2}{4}}{\mathcal{J}^7}+\frac{\frac{321 \mathcal{S}^7}{2048}+\frac{1041
   \mathcal{S}^5}{2048}+\frac{5 \mathcal{S}^3}{128}-\frac{5 \mathcal{S}}{128}}{\mathcal{J}^8}\nonumber \\
   && +\frac{-\frac{42639
   \mathcal{S}^8}{262144}-\frac{39 \mathcal{S}^6}{64}-\frac{43
   \mathcal{S}^4}{128}-\frac{\mathcal{S}^2}{4}}{\mathcal{J}^9}+\dots .
\ea
Notice that the same expansion is obtained by considering large $\J$ at fixed $\S$, i.e. 
the limit is 
fully characterized by the assumption  that $\S/\J$ is small.

We remark that  (\ref{eq:E0}) includes the contributions encoded 
by the functions $a_{10}(u) = b_{00}(u)$ and $a_{20}(u)=b_{02}(u)$ (see (\ref{ma}) and (\ref{cv})). Indeed, they can be 
computed in string theory at classical level 
and, from (\ref{sl2-charges}), it is possible to derive the following elliptic parametrizations
\ba
\label{ell1}
a_{10}(u) &=& \frac{1}{2\,\pi^{2}}\,\mathbb{K}\,\big[(\rho^{2}+1)\,\mathbb{K}-2\,\mathbb{E}\big], \\
\label{ell2}
a_{20}(u) &=& \frac{1}{8\,\pi^{4}}\,\mathbb{K}^{3}\,\big[(
-\rho^{4}+4\,\rho^{3}+2\,\rho^{2}+4\,\rho-1
)\,\mathbb{K}-8\,\rho\,\mathbb{E}\big],
\ea
where $\mathbb{E}=\mathbb{E}(1-\rho^{2})$, $\mathbb{K}=\mathbb{K}(1-\rho^{2})$, and the parameter $\rho$ is the following implicit function of the ratio $u = \S/\J$
\be
u = \frac{1}{2}\bigg(\frac{\mathbb E}{\rho\,\mathbb K}-1\bigg), 
\ee
or explicitly
\be
\rho = 1-2 \sqrt{2} \sqrt{u}+4 u-\frac{9 u^{3/2}}{2 \sqrt{2}}+u^2+\frac{25 u^{5/2}}{32 \sqrt{2}}-\frac{u^3}{2}-\frac{77
   u^{7/2}}{256 \sqrt{2}}+\frac{11 u^4}{32}+\dots.
\ee
Eqs.~(\ref{ell1},\ref{ell2}) allow to expand $a_{10}(u)$ and $a_{20}(u)$ at any desired order with 
minor effor. For instance, the first terms shown in (\ref{1}) continue as follows:
\ba
a_{10}(u) &=& \frac{u}{2}-\frac{u^2}{4}+\frac{3 u^3}{16}-\frac{21 u^4}{128}+\frac{159 u^5}{1024}-\frac{315 u^6}{2048}+\frac{321
   u^7}{2048}-\frac{42639 u^8}{262144}+\frac{716283 u^9}{4194304} \nonumber \\
   && -\frac{1514943 u^{10}}{8388608}+\frac{6433587
   u^{11}}{33554432}-\frac{54724095 u^{12}}{268435456}+\frac{465239631 u^{13}}{2147483648}+\dots, \\
a_{20}(u) &=& -\frac{u}{8}-\frac{u^2}{4}+\frac{11 u^3}{32}-\frac{27 u^4}{64}+\frac{1041 u^5}{2048}-\frac{39 u^6}{64}+\frac{11937
   u^7}{16384}-\frac{56937 u^8}{65536}+\frac{8663721 u^9}{8388608} \nonumber \\
   && -\frac{5131215 u^{10}}{4194304}+\frac{24214455
   u^{11}}{16777216}-\frac{1777563 u^{12}}{1048576}+\frac{2127741405 u^{13}}{1073741824}+\dots.
\ea

\subsection{One-loop  quantum  corrections}

The one-loop energy $e_{1}(\S, \J)$  in (\ref{stri}) 
  has been computed in the algebraic curve formalism in \cite{gssv,gva}. The calculation in 
  \cite{gva} is 
done at fixed $\J$ and small $\S$ and provides 
closed expressions for the coefficients $e_{1,n}(\J)$ appearing in 
the expansion 
\be
\label{eij}
e_{1}(\S, \J) = \sum_{n=1}^{\infty} e_{1,n}(\J)\,\S^{n}.
\ee
The large $\J$ expansion of $e_{1,n}(\J)$ is tricky because these coefficients are given in \cite{gva} 
as infinite sums. These  we may choose
 regularize by the $\zeta$-function method as
 in \cite{btz,mtt1,mtt2}.
This procedure is known to miss exponentially suppressed terms  
$\sim e^{-2\pi\J}$. This point has been already  discussed 
in \cite{SchaferNameki:2006gk}  and we shall return to this issue 
in Appendix  E. 
Explicitly, we find 
\ba
\label{check1}
e_{1,1}(\J) &=& 
-\frac{1}{2 \mathcal{J}^3}+\frac{1}{2 \mathcal{J}^5}-\frac{1}{2 \mathcal{J}^7}+\frac{1}{2
   \mathcal{J}^9}+\dots, \\
\label{check2}
e_{1,2}(\J) &=& \frac{\frac{1}{8}-\frac{\pi ^2}{12}}{\mathcal{J}^4}+\frac{\frac{5}{8}+\frac{\pi
   ^2}{24}}{\mathcal{J}^6}+\frac{-\frac{99}{64}-\frac{\pi ^2}{32}}{\mathcal{J}^8}+\frac{\frac{85}{32}+\frac{5 \pi
   ^2}{192}}{\mathcal{J}^{10}}+\frac{-\frac{4025}{1024}-\frac{35 \pi
   ^2}{1536}}{\mathcal{J}^{12}}+\dots, \\
e_{1,3}(\J) &=&\frac{\frac{3}{64}+\frac{\pi ^2}{24}+\frac{\pi ^4}{180}}{\mathcal{J}^5}+\frac{\frac{\pi
   ^2}{12}-\frac{11}{16}}{\mathcal{J}^7}+\frac{\frac{3}{16}-\frac{\pi ^2}{8}}{\mathcal{J}^9}+\frac{\frac{17}{8}+\frac{\pi
   ^2}{6}}{\mathcal{J}^{11}}+\frac{-\frac{459}{64}-\frac{5 \pi
   ^2}{24}}{\mathcal{J}^{13}}+\dots, \\
e_{1,4}(\J) &=&\frac{-\frac{99}{512}+\frac{\pi ^2}{384}-\frac{\pi ^4}{240}-\frac{\pi
   ^6}{1512}}{\mathcal{J}^6}+\frac{\frac{561}{1024}-\frac{29 \pi ^2}{256}-\frac{17 \pi ^4}{2880}-\frac{\pi
   ^6}{3024}}{\mathcal{J}^8} \nonumber \\
   && +\frac{\frac{3579}{4096}-\frac{41 \pi ^2}{3072}+\frac{\pi ^4}{288}+\frac{\pi
   ^6}{12096}}{\mathcal{J}^{10}} 
+\frac{-\frac{22303}{8192}+\frac{395 \pi ^2}{2048}-\frac{5 \pi ^4}{1536}-\frac{\pi
   ^6}{24192}}{\mathcal{J}^{12}}+\dots. 
\ea

\subsection{Large $\J$ expansion of Basso's exact slope at strong coupling}

The all-loop expression \rf{h1} 
of the slope $h_{1}(\J, \sqrt\lambda)$ defined in (\ref{sl})
was  proposed in \cite{bas} and later derived 
from the asymptotic Bethe Ansatz (ABA) in \cite{kol,bas2}.
Expanding $h_{1}(\J, \sqrt{\lambda})$ at large $\mc J$ we get the following 
ABA predictions for the $\mc O(\S)$ terms in the higher
loop string energies $e_{p}(\S, \J)$ in (\ref{stri})
\ba
\label{b-e0}
e_{0}(\S, \J) &=&  \mc S\,\Big(
\frac{1}{2 \mathcal{J}^2}-\frac{1}{8 \mathcal{J}^4}+\frac{1}{16 \mathcal{J}^6}-\frac{5}{128 \mathcal{J}^8}+\dots
\Big) + \mc O(\mc S^{2}) \\
\label{b-e1}
e_{1}(\S, \J) &=&\mc S\,\Big(
-\frac{1}{2 \mathcal{J}^3}+\frac{1}{2 \mathcal{J}^5}-\frac{1}{2 \mathcal{J}^7}+\frac{1}{2
   \mathcal{J}^9}+\dots
\Big) + \mc O(\mc S^{2}) \\
e_{2}(\S, \J) &=&\mc S\,\Big(
\frac{1}{2 \mathcal{J}^4}-\frac{11}{8 \mathcal{J}^6}+\frac{5}{2 \mathcal{J}^8}-\frac{245}{64
   \mathcal{J}^{10}}+\dots
\Big) + \mc O(\mc S^{2}) \\
e_{3}(\S, \J) &=&\mc S\,\Big(
-\frac{1}{2 \mathcal{J}^5}+\frac{13}{4 \mathcal{J}^7}-\frac{81}{8
   \mathcal{J}^9}+\frac{23}{\mathcal{J}^{11}}+\dots
   \Big) + \mc O(\mc S^{2}) \\
e_{4}(\S, \J) &=&\mc S\,\Big(
\frac{1}{2 \mathcal{J}^6}-\frac{57}{8 \mathcal{J}^8}+\frac{585}{16
   \mathcal{J}^{10}}+\dots
   \Big) + \mc O(\mc S^{2})\ , ...
\nonumber
\ea
Eqs. (\ref{b-e0}) and (\ref{b-e1}) of course  agree with the linear part of (\ref{eq:E0}) and with 
(\ref{check1}) respectively.

\section{String theory data for the $\mksu$ folded string}
\label{app:string-su2}

\subsection{Classical charges and large $\J$ limit}

The classical charges of the $\mathfrak{su}(2)$ folded string can be written in parametric form 
as the following combination of elliptic integrals depending on the complex cut endpoints $a, b$  \cite{bfst}
\ba
\mathcal J &=& -\frac{1}{2\,\pi}\frac{ab+1}{ab}\,\Big[b\,\mathbb E\Big(1-\frac{a^{2}}{b^{2}}\Big)
-a\,\mathbb K\Big(1-\frac{a^{2}}{b^{2}}\Big)\Big], \nonumber \\
\mathcal J' &=& \frac{1}{2\,\pi}\frac{ab-1}{ab}\,\Big[b\,\mathbb E\Big(1-\frac{a^{2}}{b^{2}}\Big)
+a\,\mathbb K\Big(1-\frac{a^{2}}{b^{2}}\Big)\Big], \\
\mathcal E_{0} &=& -\frac{1}{\pi}\,\frac{1}{b}\,\sqrt{(a^{2}-1)(b^{2}-1)}\,\mathbb K\Big(1-\frac{a^{2}}{b^{2}}\Big),
\nonumber
\ea
where $\E_{0}=\J+\J'+e_{0}(\J', \J)$ 
as in   (\ref{stri}) 
with the obvious
replacement $\S\to \J'$. The small $\mc J'$ expansion of $\E_{0}$ at fixed $\mc J$ reads
 \ba
\label{eq:su2-E0}
 \mc E_{0} = \mathcal{J}+\frac{\sqrt{\mathcal{J}^2+1} }{\mathcal{J}}\,\mc J'-\frac{3 \mathcal{J}^2+2 }{4 \big(\mathcal{J}^5+\mathcal{J}^3\big)}\,\mc J'^{2}+\frac{15 \mathcal{J}^6+33
   \mathcal{J}^4+28 \mathcal{J}^2+8 }{16 \mathcal{J}^5 \big(\mathcal{J}^2+1\big)^{5/2}}\,
   \mc J'^{3}+\dots 
\ea
Taking the large $\J$ limit of (\ref{eq:su2-E0}), we obtain
\ba
\label{eq:su2-classical}
\mc E_{0} &=& \mathcal{J}+\mathcal{J}'+\frac{\mathcal{J}'}{2 \mathcal{J}^2}-\frac{3 \mathcal{J}'^2}{4 \mathcal{J}^3}+\frac{\frac{15 \mathcal{J}'^3}{16}-\frac{\mathcal{J}'}{8}}{\mathcal{J}^4}+\frac{\frac{\mathcal{J}'^2}{4}-\frac{139 \mathcal{J}'^4}{128}}{\mathcal{J}^5}+\frac{\frac{1239 \mathcal{J}'^5}{1024}-\frac{9 \mathcal{J}'^3}{32}+\frac{\mathcal{J}'}{16}}{\mathcal{J}^6} \nonumber \\
   && +\frac{-\frac{2697 \mathcal{J}'^6}{2048}
   +\frac{7 \mathcal{J}'^4}{64}-\frac{\mathcal{J}'^2}{4}}{\mathcal{J}^7}+\frac{\frac{1445 \mathcal{J}'^7}{1024}+\frac{761
   \mathcal{J}'^5}{2048}+\frac{89 \mathcal{J}'^3}{128}-\frac{5 \mathcal{J}'}{128}}{\mathcal{J}^8} \nonumber \\
   && +\frac{-\frac{392049
   \mathcal{J}'^8}{262144}-\frac{81 \mathcal{J}'^6}{64}-\frac{213 \mathcal{J}'^4}{128}+\frac{\mathcal{J}'^2}{4}}{\mathcal{J}^9}+\dots 
\ea
As in the $\mksl$ folded string, the same expansion is obtained by expanding in  large $\J$ at fixed $\J'$, i.e.  the limit is 
fully characterized by the assumption 
 that the ratio $\J'/\J$ is small.

\subsection{One-loop quantum corrections}

The one-loop energy $e_{1}(\J', \J)$ has been computed in the algebraic curve approach
 in \cite{bgmrt} at
at fixed $\J$ and small $\J'$. It provides  closed expressions for the coefficients $e_{1,n}(\J)$ appearing in 
the expansion 
\be
e_{1}(\J', \J) = \sum_{n=1}^{\infty} e_{1,n}(\J)\,\J'^{n}.
\ee
Doing the same calculations as for the $\mksl$ folded string, we find
\ba
e_{1,1}(\J) &=& \frac{1}{2 \mathcal{J}^3}-\frac{1}{2 \mathcal{J}^5}+\frac{1}{2 \mathcal{J}^7}-\frac{1}{2 \mathcal{J}^9}+\frac{1}{2
   \mathcal{J}^{11}}-\frac{1}{2 \mathcal{J}^{13}}+\dots,  \\
\label{eq:su2-GV-expanded}
e_{1,2}(\J) &=& \frac{-\frac{11}{8}-\frac{\pi ^2}{12}}{\mathcal{J}^4}+\frac{\frac{15}{8}+\frac{\pi
   ^2}{24}}{\mathcal{J}^6}+\frac{-\frac{183}{64}-\frac{\pi ^2}{32}}{\mathcal{J}^8}+\frac{\frac{65}{16}+\frac{5 \pi
   ^2}{192}}{\mathcal{J}^{10}}+\frac{-\frac{5565}{1024}-\frac{35 \pi
   ^2}{1536}}{\mathcal{J}^{12}}+\dots, \\
e_{1,3}(\J) &=&\frac{\frac{157}{64}+\frac{7 \pi ^2}{24}-\frac{\pi ^4}{180}}{\mathcal{J}^5}+\frac{-\frac{67}{16}-\frac{\pi
   ^2}{12}}{\mathcal{J}^7}+\frac{\frac{157}{16}+\frac{\pi ^2}{8}}{\mathcal{J}^9}+\frac{-19-\frac{\pi
   ^2}{6}}{\mathcal{J}^{11}}+\dots. 
\ea

\section{Gauge theory data for the $\mksl$ folded string}
\label{app:gauge-sl2}

\subsection{Three-loop corrections  to anomalous dimension}

Let us  first 
compute the three-loop anomalous dimension of the ground state of the $\mksl$ spin chain, i.e.
the state dual to the spinning folded string in $AdS_3$.
The calculation is similar to the one in Appendix B of
\cite{mtt2}, but takes into account the important technical 
fact that we are interested in a state with 
highly degenerate mode numbers. An alternative derivation 
could start from the results of Appendix C.1.2 of \cite{bds}.
 The all-order Bethe equations \cite{stau1}
are written in terms of the auxiliary functions
\be
x(u) = \frac{u}{2}\,\Big(1+\sqrt{1-\frac{\lambda}{4\,\pi^{2}\,u^{2}}}\Big), \qquad 
x^{\pm}(u) = x\big(u\pm\frac{i}{2}\big),
\ee
where $u$ is the rapidity of Bethe roots.
Let us consider an even number $S$ of magnons. A generic state will be specified by the $S$ Bethe roots
$\{U_{n}(\lambda)\}_{n=1,\dots, S}$ obeying the  Bethe Ansatz equations\foot{Here we are going to 
consider  only the 3-loop corrections so the dressing phase does not contribute \ci{bes}.}
\be
\label{eq:BA}
J\,\ln\frac{x^{+}_{n}}{x^{-}_{n}}-\sum_{m\neq n}^{S}\,\ln
\Big(
\frac{x^{-}_{n}-x^{+}_{m}}{x^{+}_{n}-x^{-}_{m}}\,
\frac{1-\frac{\lambda}{16\,\pi^{2}}\,\frac{1}{x^{+}_{n}\,x^{-}_{m}}}
{1-\frac{\lambda}{16\,\pi^{2}}\,\frac{1}{x^{-}_{n}\,x^{+}_{m}}}
\Big) = 2\,\pi\,i\,N_{n}\ , \qquad n = 1, \dots, S,
\ee
where $x_{n}^{\pm} = x(U_{n})^{\pm}$. Given the Bethe roots $U_{n}(\lambda)$, the anomalous dimension $\gamma(S,J,\lambda)$ in (\ref{dima}) is given by 
\be
\gamma(S,J,\lambda) = \sum_{\ell=1}^{\infty}\lambda^{\ell}\,\gamma_{\ell}(S,J)
= \frac{\lambda}{8\,\pi^{2}} \sum_{n=1}^{S}\Big(\frac{i}{x^{+}_{n}}-\frac{i}{x^{-}_{n}}\Big).
\ee
We will be interested in the ground state of the spin chain that is characterized by a set of Bethe roots even under $U \to -U$
\be
U_{n} = (u_{1}, \dots, u_{\frac{S}{2}}, -u_{1}, \dots, -u_{\frac{S}{2}}), \qquad\quad  i=1, \dots, S.
\ee
The independent variables are thus $\{u_{n}\}_{n=1,\dots, \frac{S}{2}}$. They can be found by solving 
(\ref{eq:BA}) with $n=1, \dots, \frac{S}{2}$ and choosing the mode numbers to be equal $N_{n}=1$ 
in this range (they are $-1$  for the remaining 
 Bethe roots).

The large $J$ expansion of the Bethe roots has been worked out in Appendix B.1 of \cite{mtt2}
for the case where all $N_{n}$ are distinct. In the present case, it turns out to have the form 
\ba
\label{eq:BetheRoots}
u_{n}(J, \lambda) &=& \frac{J}{2\,\pi}+\frac{u^{(0)}_{0, n}}{\pi}\,\sqrt J+u^{(0)}_{1,n}+u^{(0)}_{2,n}\,\frac{1}{\sqrt J}
+u^{(0)}_{3,n}\,\frac{1}{J}+\dots \nonumber \\
&& \qquad + \lambda\,\Big(
\frac{u^{(1)}_{0,n}}{J}+\frac{u^{(1)}_{1,n}}{J^{3/2}}+\frac{u^{(1)}_{2,n}}{J^{2}}+\dots
\Big)+\mc O(\lambda^{2}). 
\ea
The only non-trivial problem is the determination of the constants $u^{(0)}_{0,n}$. Indeed, all the other constants
are iteratively determined by solving linear problems. Instead, the equations for $z_{n} = u^{(0)}_{0,n}$ are non linear
and read
\be
\sum_{m\neq n}^{\frac{S}{2}}\frac{1}{z_{n}-z_{m}} = 2\,z_{n},\qquad \qquad n = 1, \dots, \frac{S}{2}.
\ee
Remarkably, the solution to these equations is any permutation of the $\frac{S}{2}$ roots of the Hermite
polynomials (see for instance \cite{gv2})
\be
H_{\frac{S}{2}}(\sqrt 2\,z_{n})=0.
\ee
Working out  perturbation series for various values of $S$ we easily determine the exact 
$S$-dependence of  various $1/J^{n}$ corrections to the anomalous dimension.
The results for the one, two, and three-loop corrections to the anomalous dimension are
\ba
\label{eq:sl2-E1-gauge}
\gamma_{1}(S,J) &=& \frac{S}{2\,J^{2}}-\Big(\frac{S^{2}}{4}+\frac{S}{2}\Big)\,\frac{1}{ J^{3}}
+\Big[
\frac{3 S^3}{16}+\Big(\frac{1}{8}-\frac{\pi ^2}{12}\Big) S^2+\frac{S}{2}
\Big]\,\frac{1}{ J^{4}} \nonumber\\
&& +\Big[
-\frac{21 S^4}{128}+\Big(\frac{3}{64}+\frac{\pi ^2}{24}+\frac{\pi ^4}{180}\Big) S^3+\Big(-\frac{3}{16}+\frac{\pi ^2}{4}-\frac{\pi ^4}{90}\Big) S^2-\frac{S}{2}
\Big]\,\frac{1}{ J^{5}} \nonumber \\
&& +\Big[
\frac{159 S^5}{1024}+\Big(-\frac{99}{512}+\frac{\pi ^2}{384}-\frac{\pi ^4}{240}-\frac{\pi ^6}{1512}\Big) S^4+\Big(\frac{3}{16}-\frac{43 \pi ^2}{192}-\frac{\pi ^4}{120}+\frac{11 \pi ^6}{3780}\Big)
   S^3 \nonumber \\
 &&\ \  +\Big(\frac{5}{32}-\frac{19 \pi ^2}{48}+\frac{2 \pi ^4}{45}-\frac{\pi ^6}{315}\Big) S^2+\frac{S}{2}
\Big]\,\frac{1}{J^{6}}+\dots,  \\
\label{eq:sl2-E2-gauge}
\gamma_{2}(S,J) &=& -\frac{S}{8\,J^{4}}+\Big(\frac{-S^{2}}{4}+\frac{S}{2}\Big)\,\frac{1}{ J^{5}}
+\Big[
\frac{11 S^3}{32}+\Big(\frac{5}{8}+\frac{\pi ^2}{24}\Big) S^2-\frac{11 S}{8}
\Big]\,\frac{1}{ J^{6}} \nonumber\\
&& +\Big[
-\frac{27 S^4}{64}+\Big(\frac{\pi ^2}{12}-\frac{11}{16}\Big) S^3-\frac{17 S^2}{16}+\frac{13 S}{4}
\Big]\,\frac{1}{ J^{7}} \nonumber \\
&& +\Big[
\frac{1041 S^5}{2048}+\Big(\frac{561}{1024}-\frac{29 \pi ^2}{256}-\frac{17 \pi ^4}{2880}-\frac{\pi ^6}{3024}\Big) S^4\nonumber \\
&& \ \ \ +\Big(\frac{295}{256}-\frac{131 \pi ^2}{384}-\frac{\pi ^4}{144}+\frac{11 \pi ^6}{7560}\Big)
   S^3 \nonumber\\
&& \ \ \ +\Big(\frac{25}{16}-\frac{29 \pi ^2}{32}+\frac{\pi ^4}{40}-\frac{\pi ^6}{630}\Big) S^2-\frac{57 S}{8}
\Big]\,\frac{1}{J^{8}}+\dots,  \\
\label{eq:sl2-E3-gauge}
\gamma_{3}(S,J) &=& \frac{S}{16\,J^{6}}+\Big(\frac{3\,S^{2}}{8}-\frac{S}{2}\Big)\,\frac{1}{ J^{7}}
+\Big[
\frac{S^3}{128}+\Big(-\frac{151}{64}-\frac{\pi ^2}{32}\Big) S^2+\frac{5 S}{2}
\Big]\,\frac{1}{ J^{8}}\nonumber \\
&& +\Big[
-\frac{45 S^4}{128}+\Big(\frac{7}{32}-\frac{\pi ^2}{4}\Big) S^3  +\frac{305 S^2}{32}-\frac{81 S}{8}
\Big]\,\frac{1}{ J^{9}}\nonumber \\
&& +\Big[
\frac{5949 S^5}{8192}+\Big(\frac{5219}{4096}+\frac{119 \pi ^2}{3072}+\frac{\pi ^4}{90}+\frac{\pi ^6}{12096}\Big) S^4\nonumber \\
&& \ \ \ +\Big(-\frac{941}{512}+\frac{2059 \pi ^2}{1536}-\frac{37 \pi ^4}{2880}-\frac{11 \pi
   ^6}{30240}\Big) S^3 \nonumber \\
&& \ \ \  +\Big(-\frac{8211}{256}+\frac{165 \pi ^2}{128}-\frac{\pi ^4}{240}+\frac{\pi ^6}{2520}\Big) S^2+\frac{585 S}{16}
\Big]\,\frac{1}{J^{10}}+\dots . 
\ea

\subsection{Leading 
dressing phase  
contribution to 4-loop anomalous dimension}

The dressing phase \cite{bes} starts contributing at order $\lambda^{4}$. It is included as 
 $e^{2\,i\,\vartheta_{nm}}$  under the $\ln$ in the Bethe equations (\ref{eq:BA}).
The leading contribution to $\vartheta_{nm}$ 
 can be written as 
\be
\theta_{nm} = 4\,\zeta({3})\,\Big(\frac{\lambda}{16\pi^{2}}\Big)^{3}\,\Big[
Q_{2}(u_{n})\,Q_{3}(u_{m})-Q_{2}(u_{m})\,Q_{3}(u_{n})\Big]+ O(\lambda^4)
\ee
where the higher charges $Q_{r}(u)$ are
\be
Q_{r}(u) = \frac{i}{r-1}\Big[
\frac{1}{(u+i/2)^{r-1}}-\frac{1}{(u-i/2)^{r-1}}
\Big].
\ee
The first few terms of the large $J$ expansion of the 4-loop  anomalous dimension
$\gamma_{4}(S,J)$ are found to be
%
\ba
\label{lo-dressing}
\lefteqn{\gamma_{4}(S, J) =  -\frac{\zeta(3)}{32\,\pi^{2}}\,\frac{S^{2}}{J^{7}}
+\Big[
-\frac{5}{128}\,S+\frac{\zeta(3)}{\pi^{2}}\,\Big(\frac{13}{64}\,S^{2}+\frac{1}{128}\,S^{3}\Big)
\Big]\,\frac{1}{J^{8}}} &&  \nonumber \\
&&
\qquad +\Big\{
\frac{1}{2}S-\frac{7}{16}\,S^{2}+\frac{\zeta(3)}{\pi^{2}}\,\Big[
-\frac{7}{8}\,S^{2}-\Big(\frac{1}{128}-\frac{\pi^{2}}{32}\Big)\,S^{3}+\frac{1}{256}S^{4}\Big]
\Big\}\,\frac{1}{J^{9}}+\dots~
\ea
Note, in particular, 
 that  the linear terms in $S$ here match the corresponding terms 
in the strong-coupling expressions \rf{b-e0},\rf{b-e1} and that 
 dressing phase contributions start with $S^2$ terms.

\section{Gauge theory data for the $\mksu$ folded string}
\label{app:gauge-su2}

The corresponding 
calculation in the $\mksu$ sector is completely similar to the one in the $\mathfrak{sl}(2)$ sector. 
Let $L = J+J'$ denote 
the length of the $\mathfrak{su}(2)$ spin chain. The ground state has $J'$ magnons and its 
three loop anomalous dimension turns out to be\foot{The notation is again that of (\ref{dima}) with the obvious replacement $S\to J'$ and with $L$ playing here the role that  $J$ had in the $\mksl$ sector.}
\ba
\gamma_{1}(J',L) &=& \frac{J'}{2\,L^{2}}+\Big(\frac{J'^{2}}{4}+\frac{J'}{2}\Big)\,\frac{1}{ L^{3}}
+\Big[
\frac{3 J'^3}{16}+\Big(\frac{1}{8}-\frac{\pi ^2}{12}\Big) J'^2+\frac{J'}{2}
\Big]\,\frac{1}{ L^{4}}\nonumber\\
&& +\Big[
\frac{21 J'^4}{128}+\Big(-\frac{3}{64}-\frac{\pi ^2}{24}-\frac{\pi ^4}{180}\Big) J'^3+\Big(\frac{3}{16}-\frac{\pi
   ^2}{4}+\frac{\pi ^4}{90}\Big) J'^2+\frac{J'}{2}
\Big]\,\frac{1}{ L^{5}} \nonumber \\
&& +\Big[
\frac{159 J'^5}{1024}+\Big(-\frac{99}{512}+\frac{\pi ^2}{384}-\frac{\pi ^4}{240}-\frac{\pi ^6}{1512}\Big) J'^4+\Big(\frac{3}{16}-\frac{43 \pi ^2}{192}-\frac{\pi ^4}{120}+\frac{11 \pi ^6}{3780}\Big)
   J'^3 \nonumber \\
 && \ \ \ +\Big(\frac{5}{32}-\frac{19 \pi ^2}{48}+\frac{2 \pi ^4}{45}-\frac{\pi ^6}{315}\Big) J'^2+\frac{J'}{2}
\Big]\,\frac{1}{L^{6}}+\dots,  \\
\gamma_{2}(J',L) &=& -\frac{J'}{8\,L^{4}}+\Big(\frac{-J'^{2}}{4}-\frac{J'}{2}\Big)\,\frac{1}{ L^{5}}
+\Big[
-\frac{9 J'^3}{32}+\Big(\frac{\pi ^2}{24}-\frac{5}{8}\Big) J'^2-\frac{11 J'}{8}\Big]\,\frac{1}{ L^{6}}\nonumber\\
&& +\Big[
-\frac{21 J'^4}{32}+\Big(\frac{\pi ^2}{3}-\frac{7}{8}\Big) J'^3+\Big(\frac{2 \pi ^2}{3}-\frac{19}{8}\Big) J'^2-\frac{13
   J'}{2}
\Big]\,\frac{1}{ L^{7}} \nonumber \\
&& +\Big[
-\frac{807 J'^5}{2048}+\Big(-\frac{167}{1024}+\frac{137 \pi ^2}{768}+\frac{13 \pi ^4}{960}-\frac{\pi ^6}{3024}\Big)
   J'^4\nonumber \\
   &&\ \ \ +\Big(-\frac{209}{256}+\frac{349 \pi ^2}{384}-\frac{11 \pi ^4}{240}+\frac{11 \pi ^6}{7560}\Big)
   J'^3 \nonumber \\
   &&\ \ \ +\Big(-\frac{31}{16}+\frac{121 \pi ^2}{96}+\frac{\pi ^4}{40}-\frac{\pi ^6}{630}\Big) J'^2-\frac{57 J'}{8}
\Big]\,\frac{1}{L^{8}}+\dots,  \\
\gamma_{3}(J',L) &=& \frac{J'}{16\,L^{6}}+\Big(\frac{J'^{2}}{4}+\frac{J'}{2}\Big)\,\frac{1}{ L^{7}}
+\Big[
\frac{37 J'^3}{128}+\Big(\frac{93}{64}-\frac{\pi ^2}{32}\Big) J'^2+\frac{5 J'}{2}
\Big]\,\frac{1}{ L^{8}}\nonumber\\
&& +\Big[
\frac{49 J'^4}{128}+\Big(\frac{31}{32}-\frac{\pi ^2}{4}\Big) J'^3+\Big(\frac{175}{32}-\frac{\pi ^2}{2}\Big) J'^2+\frac{81
   J'}{8}
\Big]\,\frac{1}{ L^{9}}\nonumber \\
&& +\Big[
\frac{4437 J'^5}{8192}+\Big(\frac{2299}{4096}-\frac{403 \pi ^2}{1024}-\frac{\pi ^4}{96}+\frac{\pi ^6}{12096}\Big)
   J'^4\nonumber\\
   && \ \ \  +\Big(\frac{1031}{512}-\frac{1047 \pi ^2}{512}+\frac{29 \pi ^4}{960}-\frac{11 \pi ^6}{30240}\Big)
   J'^3 \nonumber \\
   &&\ \ \ +\Big(\frac{4365}{256}-\frac{475 \pi ^2}{128}-\frac{\pi ^4}{240}+\frac{\pi ^6}{2520}\Big) J'^2+\frac{585 J'}{16}
\Big]\,\frac{1}{L^{10}}+\dots . 
\ea
A non trivial check of these expression is the equality of the dimensions 
of the 2-magnon states in the 
$\mathfrak{sl}(2)$ and $\mathfrak{su}(2)$ sectors
implied by  superconformal invariance (see for instance \cite{bm})
\be
\underbrace{\gamma_{\ell}(J'=2, L=2+J) }_{\mathfrak{su}(2)}= \underbrace{\gamma_{\ell}(S=2, J)}_{\mathfrak{sl}(2)} \ .
\ee
The first two terms in $\gamma_1$  were   first computed in   
\cite{Minahan:2002ve}.

In order to compare with string theory, we are to  take into account that $L = J+J'$ and re-expand  
at large $J$. The resulting expressions read~\footnote{With an  abuse of the notation we 
do not distinguish here  between 
$\gamma_{\ell}(J', J+J')$ and $ \gamma_{\ell}(J', J)$.}
\ba
\label{eq:su2-E1-gauge}
\gamma_{1}(J',J) &=& 
\frac{J'}{2 J^2}+\Big(\frac{J'}{2}-\frac{3 J'^2}{4}\Big)\,\frac{1}{J^3}+\Big[\frac{15 J'^3}{16}+\Big(-\frac{11}{8}-\frac{\pi ^2}{12}\Big)
   J'^2+\frac{J'}{2}\Big]\,\frac{1}{J^4} \nonumber\\
&& +\Big[-\frac{139 J'^4}{128}+\Big(\frac{157}{64}+\frac{7 \pi ^2}{24}-\frac{\pi ^4}{180}\Big)
   J'^3+\Big(-\frac{29}{16}-\frac{\pi ^2}{4}+\frac{\pi ^4}{90}\Big) J'^2+\frac{J'}{2}\Big]\,\frac{1}{J^5} \nonumber \\
   &&+\Big[\frac{1239
   J'^5}{1024}+\Big(-\frac{1899}{512}-\frac{239 \pi ^2}{384}+\frac{17 \pi ^4}{720}-\frac{\pi ^6}{1512}\Big)
   J'^4\nonumber \\
   &&\ \ \ +\Big(\frac{17}{4}+\frac{197 \pi ^2}{192}-\frac{23 \pi ^4}{360}+\frac{11 \pi ^6}{3780}\Big)
   J'^3 \nonumber \\
   &&\ \ \ +\Big(-\frac{75}{32}-\frac{19 \pi ^2}{48}+\frac{2 \pi ^4}{45}-\frac{\pi ^6}{315}\Big)
   J'^2+\frac{J'}{2}\Big]\,\frac{1}{J^6}+\dots, 
 \\
\label{eq:su2-E2-gauge}
\gamma_{2}(J',J) &=& 
-\frac{J'}{8 J^4}+\Big(\frac{J'^2}{4}-\frac{J'}{2}\Big)\,\frac{1}{J^5}+\Big[-\frac{9 J'^3}{32}+\Big(\frac{15}{8}+\frac{\pi ^2}{24}\Big)
   J'^2-\frac{11 J'}{8}\Big]\,\frac{1}{J^6} \nonumber\\
   &&+\Big[\frac{7 J'^4}{64}+\Big(-\frac{67}{16}-\frac{\pi ^2}{12}\Big)
   J'^3+\Big(\frac{113}{16}+\frac{\pi ^2}{3}\Big) J'^2-\frac{13 J'}{4}\Big]\,\frac{1}{J^7} \nonumber \\
   &&+\Big[\frac{761
   J'^5}{2048}+\Big(\frac{7449}{1024}-\frac{29 \pi ^2}{256}+\frac{13 \pi ^4}{960}-\frac{\pi ^6}{3024}\Big)
   J'^4\nonumber \\
   && \ \ \ +\Big(-\frac{5473}{256}-\frac{547 \pi ^2}{384}-\frac{11 \pi ^4}{240}+\frac{11 \pi ^6}{7560}\Big)
   J'^3 \nonumber \\
   &&\ \ \ +\Big(\frac{333}{16}+\frac{121 \pi ^2}{96}+\frac{\pi ^4}{40}-\frac{\pi ^6}{630}\Big) J'^2-\frac{57
   J'}{8}\Big]\,\frac{1}{J^8}+\dots, \\
\label{eq:su2-E3-gauge}
\gamma_{3}(J',J) &=&  
\frac{J'}{16 J^6}+\Big(\frac{J'}{2}-\frac{J'^2}{8}\Big)\,\frac{1}{J^7}+\Big[-\frac{19 J'^3}{128}+\Big(-\frac{131}{64}-\frac{\pi
   ^2}{32}\Big) J'^2+\frac{5 J'}{2}\Big]\frac{1}{J^8} \nonumber\\
   && +\Big[\frac{201 J'^4}{128}+\frac{107 J'^3}{32}+\Big(-\frac{465}{32}-\frac{\pi
   ^2}{2}\Big) J'^2+\frac{81 J'}{8}\Big]\frac{1}{J^9} \nonumber \\
   &&+\Big[-\frac{46059 J'^5}{8192}+\Big(\frac{8827}{4096}+\frac{749 \pi
   ^2}{1024}-\frac{\pi ^4}{96}+\frac{\pi ^6}{12096}\Big) J'^4 \nonumber \\
   &&+\Big(\frac{21911}{512}+\frac{1257 \pi ^2}{512}+\frac{29 \pi
   ^4}{960}-\frac{11 \pi ^6}{30240}\Big) J'^3\nonumber \\
   &&+\Big(-\frac{18963}{256}-\frac{475 \pi ^2}{128}-\frac{\pi ^4}{240}+\frac{\pi
   ^6}{2520}\Big) J'^2+\frac{585 J'}{16}\Big]\,\frac{1}{J^{10}}+\dots\ 
\ea

\section{Details of  large $\J$ expansion of the one-loop string correction}

The coefficient $e_{1,2}(\J)$ defined (\ref{eij}) is given by the exact expression \cite{gva}
\ba
\label{eq:E12}
e_{1,2}(\mc J) &=&\frac{3 \mathcal{J}^4+11 \mathcal{J}^2+17}{16 \mathcal{J}^3 \big(\mathcal{J}^2+1\big)^{5/2}}
-\sum_{n=2}^{\infty}\frac{n^2 \big(\mathcal{J}^2+2 n^2-1\big)}{\mathcal{J}^3 \big(n^2-1\big)^2 \big(\mathcal{J}^2+n^2\big)^{3/2}}\ .
\ea
Let us discuss in detail
 its large $\J$ expansion. The aim 
  will be to clarify the role
of $\zeta$-function regularization with respect to non-analytic and exponentially suppressed contributions.
A naive expansion gives
\be
e_{1,2}(\mc J) = 
\Big(
\frac{3}{16}-\sum_{n=2}^{\infty} \frac{n^{2}}{(n^{2}-1)^{2}}
\Big)\,\frac{1}{\mc J^{4}}+
\Big(
\frac{7}{32}+\sum_{n=2}^{\infty}\frac{2n^{2}-n^{4}}{2\,(n^{2}-1)^{2}}
\Big)\,\frac{1}{\mc J^{6}}+\dots
\ee
The sum in the first term  is finite and gives
the following coefficient of $1\ov \J^4$: $
\frac{3}{16}-(\frac{1}{16}+\frac{\pi^{2}}{12}) = \frac{1}{8}-\frac{\pi^{2}}{12}.$
The sum in the second  $ 1\ov \mc J^{6}$ term is divergent. We regularize it by the 
$\zeta$-function  as follows
\ba
\frac{7}{32}+\sum_{n=2}^{\infty}\frac{2n^{2}-n^{4}}{2\,(n^{2}-1)^{2}} &=& \frac{7}{32}+
\sum_{n=2}^{\infty}\Big(\frac{1}{2\,(n^{2}-1)^{2}}-\frac{1}{2}\Big) = \nonumber\\
&=& \frac{7}{32}+
\frac{\pi^{2}}{24}-\frac{11}{32}-\frac{1}{2}\big(\zeta(0)-1\big) = \frac{5}{8}+\frac{\pi^{2}}{24}\ . 
\ea
Thus the  $\zeta$-function regularization provides the following result
\be
e_{1,2}(\mc J) = \Big(\frac{1}{8}-\frac{\pi^{2}}{12}\Big)\,\frac{1}{\mc J^{4}}+
\Big(
\frac{5}{8}+\frac{\pi^{2}}{24}
\Big)\,\frac{1}{\mc J^{6}}+\dots,
\ee
The details of the $\zeta$-function regularization
 in the general  are as follows. For a rational function $R(n^{2})$ we write
\be
R(n^{2}) = \sum_{k=0}^{p} c_{k}\,n^{2k}+\mc O({1\ov n^{2}}). 
\ee
Then, our definition for the regularized sum is
\be
\sum_{n=2}^{\infty} R(n^{2}) \stackrel{\zeta-  reg.}{=} \sum_{n=2}^{\infty} \Big[R(n^{2}) -\sum_{k=0}^{p} c_{k}\,n^{2k}\Big]
+\sum_{k=0}^{p} c_{k}\,\Big(\zeta(-2k)-1\Big).
\ee
The above procedure misses  non-analytic terms with odd powers of $1\ov \mc J$. These are due to the 
dressing phase in the all-order  Bethe ansatz equations~\cite{bes} and are 
not captured by the $\zeta$-function regularization. They  affect the coefficients of $\mc S^{n}$ terms starting
from $n=2$.

In order to find them, at least at one-loop order, one has to compute the infinite sums in 
$e_{1,n}(\J)$ exactly at finite $\mc J$ and 
then perform  the large $\mc J$ expansion. In the 
specific case of $e_{1,2}(\J)$, it can be written  as 
\be
e_{1,2}(\mc J) = e_{1,2}^{\rm anomaly}(\mc J) + e_{1,2}^{\rm dressing}(\mc J) +
e_{1,2}^{\rm wrapping}(\mc J),
\ee
where we have used the terminology of \cite{gssv,gva} and  have split the correction into 
the so-called {\em anomaly} term, the {\em dressing} phase contribution, and the {\em wrapping} contribution. The explicit expressions for the first two can be found in Appendix A of \cite{gva}:
\ba 
e_{1,2}^{\rm anomaly}(\mc J)  &=& \frac{2 \mathcal{J}^4+15 \mathcal{J}^2+4}{16 \mathcal{J}^3 \big(\mathcal{J}^2+1\big)^{5/2}}-\frac{\pi ^2}{12 \mathcal{J}^3 \sqrt{\mathcal{J}^2+1}}\ , \\
e_{1,2}^{\rm dressing}(\mc J)  &=&
\frac{\big(\mathcal{J}^2+2\big) \coth ^{-1}\big(\sqrt{\mathcal{J}^2+1}\big)-\sqrt{\mathcal{J}^2+1}}{2 \mathcal{J}^3
   \big(\mathcal{J}^2+1\big)^{3/2}}\ .
\ea
We have computed the wrapping contribution following 
  \cite{SchaferNameki:2006gk}:
\ba
&&e_{1,2}^{\rm wrapping}(\mc J) = \int_{\mc J}^{\infty} dt\,
\frac{t^{2}}{\mathcal{J}^5 \big(t^2+1\big)^3 \sqrt{t^{2}-\mathcal{J}^{2}}}\Big[
2 \Big(-\mathcal{J}^2+\big(\mathcal{J}^2+3\big) t^2+1\Big) \big[\coth (\pi  t)-1\big]
\nonumber\\
&&  \qquad\qquad -\pi  t \big(t^2+1\big) \big(-\mathcal{J}^2+2
   t^2+1\big) \text{csch}^2(\pi  t)\Big],
\ea
confirming that at large $\mc J$ it is suppressed as $\mc O(e^{-2\,\pi\,\mc J})$.
The expansion of $e_{1,2}^{\rm anomaly}(\mc J)$ is a regular power series in $1/\mc J^{2}$
\be
e_{1,2}^{\rm anomaly}(\mc J) = \frac{\frac{1}{8}-\frac{\pi ^2}{12}}{\mathcal{J}^4}+\frac{\frac{5}{8}+\frac{\pi ^2}{24}}{\mathcal{J}^6}+\frac{-\frac{99}{64}-\frac{\pi
   ^2}{32}}{\mathcal{J}^8}+\frac{\frac{85}{32}+\frac{5 \pi ^2}{192}}{\mathcal{J}^{10}}+\dots,
\ee
in agreement with the  previous expression (\ref{check2}).
At the same time, the expansion of $e_{1,2}^{\rm dressing}(\mc J)$ is a regular power series containing only  odd powers $1/ \mc J$
\be
e_{1,2}^{\rm dressing}(\mc J) = \frac{2}{3 \mathcal{J}^7}-\frac{16}{15 \mathcal{J}^9}+\frac{48}{35 \mathcal{J}^{11}}-\frac{512}{315 \mathcal{J}^{13}}+\frac{1280}{693
   \mathcal{J}^{15}}+\dots \ .
\ee


\end{document}